\documentclass[prd,showkeys,twocolumn,amsmath,amssymb]{revtex4}
\usepackage{graphicx}
\usepackage{epstopdf}
\usepackage{subfigure}
\usepackage{bm}
\usepackage[colorlinks,citecolor=blue,anchorcolor=red,menucolor=red,linkcolor=red,filecolor=red,runcolor=red,urlcolor=blue,frenchlinks=red]{hyperref}
\usepackage{ulem}
\usepackage{color}
\usepackage{multirow}

\allowdisplaybreaks[3]
\begin{document}
\title{Fully-strange tetraquark states with the exotic quantum numbers $J^{PC} = 0^{+-}$ and $2^{+-}$}
%

\author{Hong-Zhou Xi$^1$}
\author{Yi-Wei Jiang$^1$}
\author{Hua-Xing Chen$^1$}
\email{hxchen@seu.edu.cn}
\author{Atsushi Hosaka$^{2,3}$}
\email{hosaka@rcnp.osaka-u.ac.jp}
\author{Niu Su$^{1,2}$}
\email{suniu@seu.edu.cn}

\affiliation{$^1$School of Physics, Southeast University, Nanjing 210094, China\\
$^2$Research Center for Nuclear Physics (RCNP), Osaka University, Ibaraki 567-0047, Japan \\
$^3$Advanced Science Research Center, Japan Atomic Energy Agency (JAEA), Tokai 319-1195, Japan}

\begin{abstract}
We study the fully-strange tetraquark states with the exotic quantum numbers $J^{PC} = 0^{+-}$ and $2^{+-}$. We construct their corresponding diquark-antidiquark interpolating currents, and apply the QCD sum rule method to calculate both their diagonal and off-diagonal correlation functions. The obtained results are used to construct some mixing currents that are nearly non-correlated, from which we extract the masses of the lowest-lying states to be $M_{0^{+-}} = 2.47^{+0.33}_{-0.44}$~GeV and $M_{2^{+-}} = 3.07^{+0.25}_{-0.33}$~GeV. We apply the Fierz rearrangement to transform the diquark-antidiquark currents to be the combinations of meson-meson currents, and the obtained Fierz identities indicate that these two states may be searched for in the $P$-wave $\phi(1020) f_0(1710)/\phi(1020) f_2^\prime(1525) (\to \phi K \bar K / \phi \pi \pi)$ channels.
\end{abstract}
\maketitle
\pagenumbering{arabic}

\section{Introduction}
\label{sec:introduction}

Many candidates of exotic hadrons were observed in the past twenty years, which can not be well explained in the traditional quark model as the $\bar q q$ mesons and $qqq$ baryons~\cite{pdg}. However, most of them still have the ``traditional'' quantum numbers that the traditional hadrons can reach, making them not easy to be clearly identified as exotic hadrons. Interestingly, there exist some ``exotic'' quantum numbers that the traditional hadrons can not reach, {\it e.g.}, the spin-parity quantum numbers $J^{PC} = 0^{\pm-}/1^{-+}/2^{+-}/3^{-+}/4^{+-}/\cdots$~\cite{Chen:2022asf,Klempt:2007cp,Meyer:2015eta,Amsler:2004ps,Bugg:2004xu,Meyer:2010ku,Briceno:2017max,COMPASS:2018uzl,JPAC:2018zyd,Ketzer:2019wmd,Jin:2021vct,Meng:2022ozq,Albuquerque:2023bex}.

The hadrons with these exotic quantum numbers are definitely exotic hadrons, and their possible interpretations are compact multiquark states~\cite{Jiao:2009ra,Huang:2016rro,LEE:2020eif,Du:2012pn,Fu:2018ngx,Azizi:2019ecm,Su:2022eun}, hadronic molecules~\cite{Zhu:2013sca,Ji:2022blw,Zhang:2019ykd,Dong:2022cuw,Yang:2022lwq,Wan:2022xkx,Wang:2022sib,Yu:2022wtu}, glueballs~\cite{Qiao:2014vva,Tang:2015twt,Pimikov:2017bkk,Wilson:1974sk,Chen:2005mg,Mathieu:2008me,Meyer:2004gx,Gregory:2012hu,Athenodorou:2020ani}, and hybrid states~\cite{Zhang:2013rya,Huang:2014hya,Ho:2018cat,Chen:2021smz,Su:2022fqr,Su:2023jxb}, etc. Especially, the light hybrid states with the exotic quantum numbers $J^{PC} = 1^{-+}$ have been extensively studied in the literature~\cite{Barnes:1977hg,Hasenfratz:1980jv,Chanowitz:1982qj,Isgur:1983wj,Close:1994hc,Page:1998gz,Burns:2006wz,Qiu:2022ktc,Szczepaniak:2001rg,Iddir:2007dq,Guo:2007sm,Michael:1985ne,McNeile:1998cp,Juge:2002br,Lacock:1996ny,MILC:1997usn,Bernard:2003jd,Hedditch:2005zf,Dudek:2009qf,Dudek:2010wm,Dudek:2013yja,Chen:2022isv,Balitsky:1982ps,Govaerts:1983ka,Kisslinger:1995yw,Chetyrkin:2000tj,Jin:2002rw,Narison:2009vj,Huang:2016upt,Li:2021fwk,COMPASS:2009xrl}, since there are some experimental evidences on their existence~\cite{IHEP-Brussels-LosAlamos-AnnecyLAPP:1988iqi,E852:1998mbq,E852:2004gpn,BESIII:2022riz,BESIII:2022qzu}. We have also studied these exotic quantum numbers $J^{PC} = 1^{-+}$ in Refs.~\cite{Huang:2010dc,Chen:2010ic,Chen:2022qpd,Chen:2008qw,Chen:2008ne} through the QCD sum rule method. Besides, the isoscalar $D^* \bar D_2^*$ molecular state of $J^{PC} = 3^{-+}$ was predicted in Ref.~\cite{Zhu:2013sca} through the one-boson-exchange model, and a narrow hadronic state with the exotic quantum numbers $J^{PC} = 0^{--}$ was predicted in Ref.~\cite{Ji:2022blw} through the heavy quark spin symmetry.

In this paper we shall investigate the fully-strange tetraquark states with the exotic quantum numbers $J^{PC} = 0^{+-}$ and $2^{+-}$ through the method of QCD sum rules. Recently, we have applied this method to study the fully-strange tetraquark states of $J^{PC} = 0^{-+}/1^{\pm\pm}/4^{+-}$ in Refs.~\cite{Chen:2008ej,Chen:2008qw,Chen:2008ne,Chen:2018kuu,Cui:2019roq,Dong:2020okt,Dong:2022otb}. In the present study we shall explicitly add the covariant derivative operator in order to construct the fully-strange tetraquark currents of $J^{PC} = 0^{+-}$ and $2^{+-}$. We shall construct some diquark-antidiquark interpolating currents, and apply the QCD sum rule method to calculate both their diagonal and off-diagonal correlation functions. The obtained results will be used to construct some mixing currents that are nearly non-correlated, from which we shall extract the masses of the lowest-lying states to be
\begin{eqnarray}
\nonumber M_{0^{+-}} &=& 2.47^{+0.33}_{-0.44}{\rm~GeV} \, ,
\\[2mm] \nonumber M_{2^{+-}} &=& 3.07^{+0.25}_{-0.33}{\rm~GeV} \, .
\end{eqnarray}
With a large amount of $J/\psi$ sample, BESIII are carefully studying the physics happening around this energy region~\cite{BES:2003aic,BES:2005ega,BESIII:2010vwa,BESIII:2010gmv,BESIII:2016qzq,BESIII:2019wkp,BESIII:2020vtu}, and so do the Belle-II~\cite{Belle-II:2018jsg} and GlueX~\cite{Austregesilo:2018mno}. Therefore, the above fully-strange tetraquark states of $J^{PC} = 0^{+-}$ and $2^{+-}$ are potential exotic hadrons to be observed in the future experiments. The present study would provide not only complementary information for possible counter-candidates in the charm sector $cc\bar c \bar c$~\cite{LHCb:2020bwg,ATLAS:2023bft,CMS:2023owd}, but also systematic understanding of exotics in a wider flavor region.

This paper is organized as follows. In Sec.~\ref{sec:current} we construct the fully-strange tetraquark currents with the exotic quantum numbers $J^{PC} = 0^{+-}$ and $2^{+-}$. These currents are used in Sec.~\ref{sec:sumrule} to perform QCD sum rule analyses, where we calculate both their diagonal and off-diagonal correlation functions. Based on the obtained results, we use the diquark-antidiquark currents to perform numerical analyses in Sec.~\ref{sec:single}, and use their mixing currents to perform numerical analyses in Sec.~\ref{sec:mixing}. Sec.~\ref{sec:summary} is a summary.

\section{Tetraquark currents}
\label{sec:current}

In this section we construct the fully-strange tetraquark currents with the exotic quantum numbers $J^{PC} = 0^{+-}$ and $2^{+-}$. Note that these two quantum numbers can not be simply reached by using one quark field and one antiquark field, and they can not be reached by only using two quark fields and two antiquark fields neither. Actually, we need two quark fields and two antiquark fields together with one or more derivatives to reach them.

We have systematically constructed three independent diquark-antidiquark currents in Ref.~\cite{Dong:2022otb} using two quark fields and two antiquark fields together with two derivatives:
\begin{eqnarray}
\eta_{4^{+-};1}^{\alpha_1\alpha_2\alpha_3\alpha_4} &=& \epsilon^{abe} \epsilon^{cde} \times
\label{def:eta41}
\\ \nonumber &&
\mathcal{S} \Big\{ \big[s_a^T C \gamma_{\alpha_1} {\overset{\leftrightarrow}{D}}_{\alpha_3}{\overset{\leftrightarrow}{D}}_{\alpha_4} s_b \big] (\bar{s}_c \gamma_{\alpha_2} C \bar{s}_d^T)
\\ \nonumber &&
~ - (s_a^T C \gamma_{\alpha_1} s_b ) \big[\bar{s}_c \gamma_{\alpha_2} C {\overset{\leftrightarrow}{D}}_{\alpha_3} {\overset{\leftrightarrow}{D}}_{\alpha_4} \bar{s}_d^T\big]
\Big\},
\\[2mm] \eta_{4^{+-};2}^{\alpha_1\alpha_2\alpha_3\alpha_4} &=& (\delta^{ac} \delta^{bd} + \delta^{ad} \delta^{bc} ) \times
\label{def:eta42}
\\ \nonumber &&
\mathcal{S} \Big\{ \big[s_a^T C \gamma_{\alpha_1} \gamma_5 {\overset{\leftrightarrow}{D}}_{\alpha_3}{\overset{\leftrightarrow}{D}}_{\alpha_4} s_b \big] (\bar{s}_c \gamma_{\alpha_2} \gamma_5 C \bar{s}_d^T)
\\ \nonumber &&
~ - (s_a^T C \gamma_{\alpha_1} \gamma_5 s_b ) \big[\bar{s}_c \gamma_{\alpha_2} \gamma_5 C {\overset{\leftrightarrow}{D}}_{\alpha_3} {\overset{\leftrightarrow}{D}}_{\alpha_4} \bar{s}_d^T\big]
\Big\},
\\[2mm] \eta_{4^{+-};3}^{\alpha_1\alpha_2\alpha_3\alpha_4} &=& \epsilon^{abe} \epsilon^{cde} g^{\mu\nu} \times
\label{def:eta43}
\\ \nonumber &&
\mathcal{S} \Big\{ \big[s_a^T C \sigma_{\alpha_1\mu} {\overset{\leftrightarrow}{D}}_{\alpha_3}{\overset{\leftrightarrow}{D}}_{\alpha_4} s_b \big] (\bar{s}_c \sigma_{\alpha_2\nu} C \bar{s}_d^T)
\\ \nonumber &&
~ - (s_a^T C \sigma_{\alpha_1\mu} s_b ) \big[\bar{s}_c \sigma_{\alpha_2\nu} C {\overset{\leftrightarrow}{D}}_{\alpha_3} {\overset{\leftrightarrow}{D}}_{\alpha_4} \bar{s}_d^T\big]
\Big\}.
\end{eqnarray}
Here $a \cdots e$ are color indices, $\big[ A {\overset{\leftrightarrow}{D}}_\alpha B \big] = A [D_\alpha B] - [D_\alpha A] B$ with the covalent derivative $D_\alpha = \partial_\alpha + i g_s A_\alpha$, and the symbol $\mathcal{S}$ denotes symmetrization and subtracting trace terms in the set $\{\alpha_1 \cdots \alpha_4\}$.

\begin{figure}[hbtp]
\begin{center}
\includegraphics[width=0.45\textwidth]{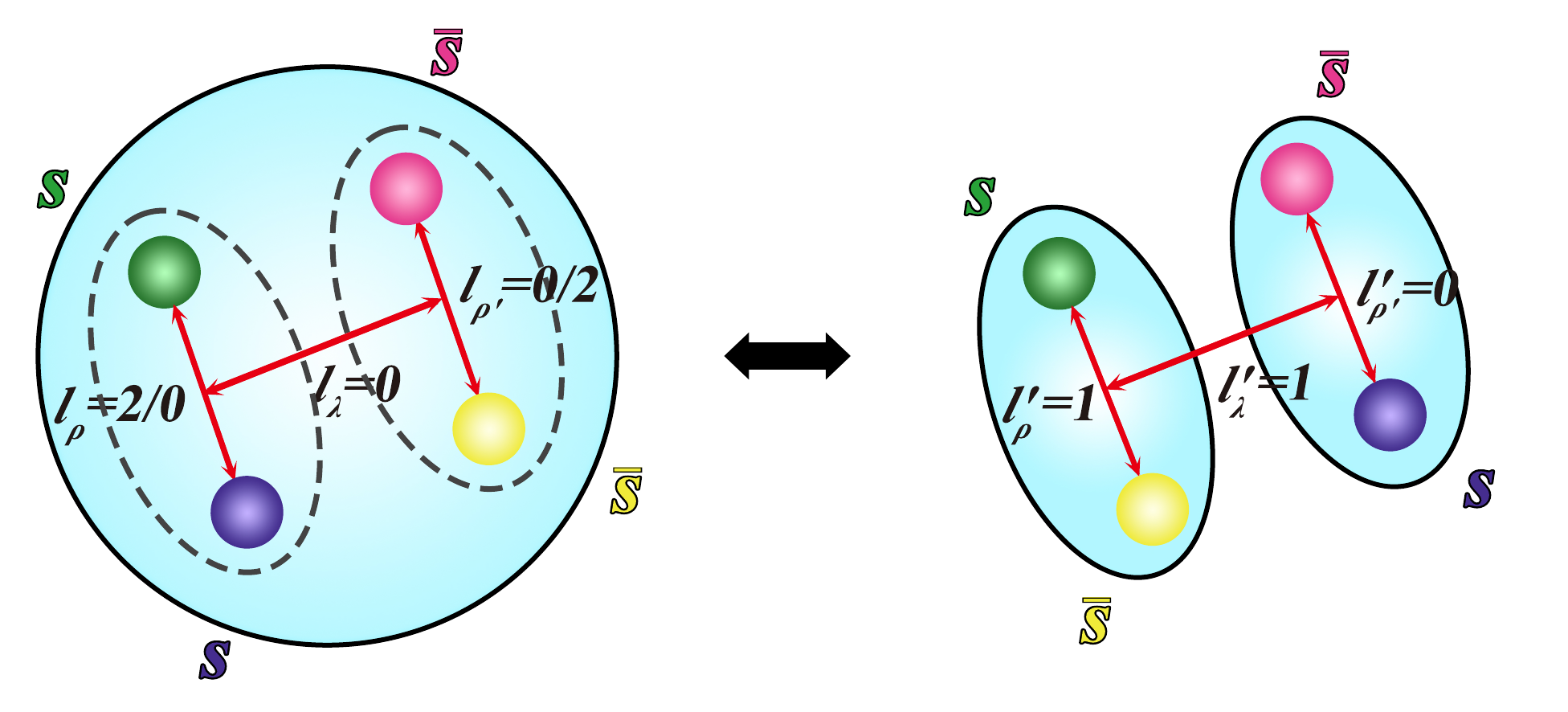}
\caption{Internal orbital angular momenta contained in the fully-strange tetraquark currents $\eta_{0/2/4^{+-};1/2/3}^{\cdots}$ and $\xi_{0/2/4^{+-};1/2/3}^{\cdots}$. We use $l_\rho$ and $l_{\rho^\prime}$ to denote the momenta inside the diquark and antidiquark respectively, and $l_\lambda$ to denote the momentum between them. We use $l^\prime_\rho$ and $l^\prime_{\rho^\prime}$ to denote the momenta inside the two mesons, and $l^\prime_\lambda$ to denote the momentum between them. The Fierz identities given in Eqs.~(\ref{eq:fierz}) indicate that the internal orbital angular momenta contained in the diquark-antidiquark currents $\eta_{0/2/4^{+-};1/2/3}^{\cdots}$, $\{L=2$; $l_\lambda = 0$, $l_\rho = 2/0$, $l_{\rho^\prime} = 0/2\}$, correspond to those contained in the meson-meson currents $\xi_{0/2/4^{+-};1/2/3}^{\cdots}$, $\{L=2$; $l_\lambda^\prime = 1$, $l_\rho^\prime = 1$, $l_{\rho^\prime}^\prime = 0\}$.}
\label{fig:relation}
\end{center}
\end{figure}

Following a similar procedure for $\eta_{4^{+-};1/2/3}^{\alpha_1\alpha_2\alpha_3\alpha_4}$, for $J^{PC} = 0^{+-}$ and $2^{+-}$ currents, we can use the spin-0 and spin-2 projection operators rather than the symmetrization operator $\mathcal{S}$:
\begin{eqnarray}
&& \mathcal{P}_{J=0}^{\alpha_1\alpha_2\alpha_3\alpha_4} = {1\over2}g_{\mu_1 \mu_3}g_{\mu_2 \mu_4}
\\ \nonumber && ~~~~~ \times \left( g^{\alpha_1 \mu_1} g^{\alpha_2 \mu_2} + g^{\alpha_1 \mu_2} g^{\alpha_2 \mu_1} - {1 \over 2} g^{\alpha_1 \alpha_2} g^{\mu_1 \mu_2} \right)
\\ \nonumber && ~~~~~ \times \left( g^{\alpha_3 \mu_3} g^{\alpha_4 \mu_4} + g^{\alpha_3 \mu_4} g^{\alpha_4 \mu_3} - {1 \over 2} g^{\alpha_3 \alpha_4} g^{\mu_3 \mu_4} \right)
\\ \nonumber &=& g^{\alpha_1 \alpha_3} g^{\alpha_2 \alpha_4} + g^{\alpha_1 \alpha_4} g^{\alpha_2 \alpha_3} - {1 \over 2} g^{\alpha_1 \alpha_2} g^{\alpha_3 \alpha_4} \, ,
\end{eqnarray}
\begin{eqnarray}
&& \mathcal{P}_{J=2;\beta_1 \beta_2}^{\alpha_1\alpha_2\alpha_3\alpha_4} = g_{\mu_2 \mu_4}
\\ \nonumber && ~~~~~ \times \left( g_{\beta_1 \mu_1} g_{\beta_2 \mu_3} + g_{\beta_1 \mu_3} g_{\beta_2 \mu_1} - {1 \over 2} g_{\beta_1 \beta_2} g_{\mu_1 \mu_3} \right)
\\ \nonumber && ~~~~~ \times \left( g^{\alpha_1 \mu_1} g^{\alpha_2 \mu_2} + g^{\alpha_1 \mu_2} g^{\alpha_2 \mu_1} - {1 \over 2} g^{\alpha_1 \alpha_2} g^{\mu_1 \mu_2} \right)
\\ \nonumber && ~~~~~ \times \left( g^{\alpha_3 \mu_3} g^{\alpha_4 \mu_4} + g^{\alpha_3 \mu_4} g^{\alpha_4 \mu_3} - {1 \over 2} g^{\alpha_3 \alpha_4} g^{\mu_3 \mu_4} \right)
\\ \nonumber &=& g^{\alpha_1 \alpha_2} g^{\alpha_3 \alpha_4} g_{\beta_1 \beta_2} - g^{\alpha_1 \alpha_2} \delta^{\alpha_3}_{\beta_1} \delta^{\alpha_4}_{\beta_2} - g^{\alpha_1 \alpha_2} \delta^{\alpha_3}_{\beta_2} \delta^{\alpha_4}_{\beta_1}
\\ \nonumber && ~~~~~~~~~~~~~~~~~~~~~ - g^{\alpha_3 \alpha_4} \delta^{\alpha_1}_{\beta_1} \delta^{\alpha_2}_{\beta_2} - g^{\alpha_3 \alpha_4} \delta^{\alpha_1}_{\beta_2} \delta^{\alpha_2}_{\beta_1}
\\ \nonumber &-& g^{\alpha_1 \alpha_3} g^{\alpha_2 \alpha_4} g_{\beta_1 \beta_2} + g^{\alpha_1 \alpha_3} \delta^{\alpha_2}_{\beta_1} \delta^{\alpha_4}_{\beta_2} + g^{\alpha_1 \alpha_3} \delta^{\alpha_2}_{\beta_2} \delta^{\alpha_4}_{\beta_1}
\\ \nonumber && ~~~~~~~~~~~~~~~~~~~~~ + g^{\alpha_2 \alpha_4} \delta^{\alpha_1}_{\beta_1} \delta^{\alpha_3}_{\beta_2} + g^{\alpha_2 \alpha_4} \delta^{\alpha_1}_{\beta_2} \delta^{\alpha_3}_{\beta_1}
\\ \nonumber &-& g^{\alpha_1 \alpha_4} g^{\alpha_2 \alpha_3} g_{\beta_1 \beta_2} - g^{\alpha_1 \alpha_4} \delta^{\alpha_2}_{\beta_1} \delta^{\alpha_3}_{\beta_2} + g^{\alpha_1 \alpha_4} \delta^{\alpha_2}_{\beta_2} \delta^{\alpha_3}_{\beta_1}
\\ \nonumber && ~~~~~~~~~~~~~~~~~~~~~ + g^{\alpha_2 \alpha_3} \delta^{\alpha_1}_{\beta_1} \delta^{\alpha_4}_{\beta_2} + g^{\alpha_2 \alpha_3} \delta^{\alpha_1}_{\beta_2} \delta^{\alpha_4}_{\beta_1} \, .
\end{eqnarray}
For $J^{PC} = 0^{+-}$, we can construct three independent diquark-antidiquark currents:
\begin{eqnarray}
\eta_{0^{+-};1} &=& \mathcal{P}_{J=0}^{\alpha_1\alpha_2\alpha_3\alpha_4} \times \epsilon^{abe} \epsilon^{cde}
\label{def:eta01}
\\ \nonumber &\times&
\big[s_a^T C \gamma_{\alpha_1} {\overset{\leftrightarrow}{D}}_{\alpha_3}{\overset{\leftrightarrow}{D}}_{\alpha_4} s_b \big] (\bar{s}_c \gamma_{\alpha_2} C \bar{s}_d^T)
\\ \nonumber &&
~ - (s_a^T C \gamma_{\alpha_1} s_b ) \big[\bar{s}_c \gamma_{\alpha_2} C {\overset{\leftrightarrow}{D}}_{\alpha_3} {\overset{\leftrightarrow}{D}}_{\alpha_4} \bar{s}_d^T\big],
\\[2mm] \eta_{0^{+-};2} &=& \mathcal{P}_{J=0}^{\alpha_1\alpha_2\alpha_3\alpha_4} \times (\delta^{ac} \delta^{bd} + \delta^{ad} \delta^{bc} )
\label{def:eta02}
\\ \nonumber &\times&
\big[s_a^T C \gamma_{\alpha_1} \gamma_5 {\overset{\leftrightarrow}{D}}_{\alpha_3}{\overset{\leftrightarrow}{D}}_{\alpha_4} s_b \big] (\bar{s}_c \gamma_{\alpha_2} \gamma_5 C \bar{s}_d^T)
\\ \nonumber &&
~ - (s_a^T C \gamma_{\alpha_1} \gamma_5 s_b ) \big[\bar{s}_c \gamma_{\alpha_2} \gamma_5 C {\overset{\leftrightarrow}{D}}_{\alpha_3} {\overset{\leftrightarrow}{D}}_{\alpha_4} \bar{s}_d^T\big],
\\[2mm] \eta_{0^{+-};3} &=& \mathcal{P}_{J=0}^{\alpha_1\alpha_2\alpha_3\alpha_4} \times \epsilon^{abe} \epsilon^{cde} g^{\mu\nu}
\label{def:eta03}
\\ \nonumber &\times&
\big[s_a^T C \sigma_{\alpha_1\mu} {\overset{\leftrightarrow}{D}}_{\alpha_3}{\overset{\leftrightarrow}{D}}_{\alpha_4} s_b \big] (\bar{s}_c \sigma_{\alpha_2\nu} C \bar{s}_d^T)
\\ \nonumber &&
~ - (s_a^T C \sigma_{\alpha_1\mu} s_b ) \big[\bar{s}_c \sigma_{\alpha_2\nu} C {\overset{\leftrightarrow}{D}}_{\alpha_3} {\overset{\leftrightarrow}{D}}_{\alpha_4} \bar{s}_d^T\big],
\end{eqnarray}
For $J^{PC} = 2^{+-}$, we can also construct three independent diquark-antidiquark currents:
\begin{eqnarray}
\eta_{2^{+-};1}^{\beta_1\beta_2} &=& \mathcal{P}_{J=2;\beta_1 \beta_2}^{\alpha_1\alpha_2\alpha_3\alpha_4} \times \epsilon^{abe} \epsilon^{cde}
\label{def:eta21}
\\ \nonumber &\times&
\big[s_a^T C \gamma_{\alpha_1} {\overset{\leftrightarrow}{D}}_{\alpha_3}{\overset{\leftrightarrow}{D}}_{\alpha_4} s_b \big] (\bar{s}_c \gamma_{\alpha_2} C \bar{s}_d^T)
\\ \nonumber &&
~ - (s_a^T C \gamma_{\alpha_1} s_b ) \big[\bar{s}_c \gamma_{\alpha_2} C {\overset{\leftrightarrow}{D}}_{\alpha_3} {\overset{\leftrightarrow}{D}}_{\alpha_4} \bar{s}_d^T\big],
\\[2mm] \eta_{2^{+-};2}^{\beta_1\beta_2} &=& \mathcal{P}_{J=2;\beta_1 \beta_2}^{\alpha_1\alpha_2\alpha_3\alpha_4} \times (\delta^{ac} \delta^{bd} + \delta^{ad} \delta^{bc} )
\label{def:eta22}
\\ \nonumber &\times&
\big[s_a^T C \gamma_{\alpha_1} \gamma_5 {\overset{\leftrightarrow}{D}}_{\alpha_3}{\overset{\leftrightarrow}{D}}_{\alpha_4} s_b \big] (\bar{s}_c \gamma_{\alpha_2} \gamma_5 C \bar{s}_d^T)
\\ \nonumber &&
~ - (s_a^T C \gamma_{\alpha_1} \gamma_5 s_b ) \big[\bar{s}_c \gamma_{\alpha_2} \gamma_5 C {\overset{\leftrightarrow}{D}}_{\alpha_3} {\overset{\leftrightarrow}{D}}_{\alpha_4} \bar{s}_d^T\big],
\\[2mm] \eta_{2^{+-};3}^{\beta_1\beta_2} &=& \mathcal{P}_{J=2;\beta_1 \beta_2}^{\alpha_1\alpha_2\alpha_3\alpha_4} \times \epsilon^{abe} \epsilon^{cde} g^{\mu\nu}
\label{def:eta23}
\\ \nonumber &\times&
\big[s_a^T C \sigma_{\alpha_1\mu} {\overset{\leftrightarrow}{D}}_{\alpha_3}{\overset{\leftrightarrow}{D}}_{\alpha_4} s_b \big] (\bar{s}_c \sigma_{\alpha_2\nu} C \bar{s}_d^T)
\\ \nonumber &&
~ - (s_a^T C \sigma_{\alpha_1\mu} s_b ) \big[\bar{s}_c \sigma_{\alpha_2\nu} C {\overset{\leftrightarrow}{D}}_{\alpha_3} {\overset{\leftrightarrow}{D}}_{\alpha_4} \bar{s}_d^T\big].
\end{eqnarray}
The internal orbital angular momenta contained in the above diquark-antidiquark currents are all
\begin{equation}
\eta_{0/2/4^{+-};1/2/3}^{\cdots} : L=2 ; \, l_\lambda = 0 , \, l_\rho = 2/0 , \, l_{\rho^\prime} = 0/2 ,
\end{equation}
where $L$ is the total orbital angular momentum, $l_\rho$ and $l_{\rho^\prime}$ are the momenta inside the diquark and antidiquark respectively, and $l_\lambda$ is the momentum between the diquark and antidiquark, as depicted in Fig.~\ref{fig:relation}.

Among the above diquark-antidiquark currents, $\eta_{0/2/4^{+-};1}^{\cdots}$ and $\eta_{0/2/4^{+-};3}^{\cdots}$ have the antisymmetric color structure $[ss]_{\mathbf{\bar 3}_C}[\bar s \bar s]_{\mathbf{3}_C}$, and $\eta_{0/2/4^{+-};2}^{\cdots}$ have the symmetric color structure $[ss]_{\mathbf{6}_C}[\bar s \bar s]_{\mathbf{\bar 6}_C}$, so the internal structure of $\eta_{0/2/4^{+-};1}^{\cdots}$ and $\eta_{0/2/4^{+-};3}^{\cdots}$ are more stable than that of $\eta_{0/2/4^{+-};2}^{\cdots}$. Moreover, the currents $\eta_{0/2/4^{+-};1}^{\cdots}$ are constructed by using the $S$-wave diquark field $s_a^T C \gamma_\mu s_b$ of $J^P=1^+$, and the currents $\eta_{0/2/4^{+-};3}^{\cdots}$ are constructed by using the diquark field $s_a^T C \sigma_{\mu\nu} s_b$ of $J^P=1^\pm$ that contains both the $S$- and $P$-wave components, so they may lead to better QCD sum rule results. Oppositely, the currents $\eta_{0/2/4^{+-};2}^{\cdots}$ are constructed by using the $P$-wave diquark field $s_a^T C \gamma_\mu \gamma_5 s_b$ of $J^P=1^-$, so their predicted masses are probably larger. The results of Ref.~\cite{Dong:2022otb} have partly verified these analyses, as summarized in Table~\ref{tab:result}.

Besides the diquark-antidiquark configuration, we can also investigate the meson-meson configuration. We have constructed three independent meson-meson currents of $J^{PC} = 4^{+-}$ in Ref.~\cite{Dong:2022otb}:
\begin{eqnarray}
\xi_{4^{+-};1}^{\alpha_1\alpha_2\alpha_3\alpha_4} &=&
\mathcal{S} \Big\{
\big[\bar s_a \gamma_{\alpha_1} {\overset{\leftrightarrow}{D}}_{\alpha_3} s_a \big] {\overset{\leftrightarrow}{D}}_{\alpha_4} (\bar{s}_b \gamma_{\alpha_2} s_b)
\Big\} ,
\label{def:xi41}
\\[2mm] \nonumber \xi_{4^{+-};2}^{\alpha_1\alpha_2\alpha_3\alpha_4} &=&
\mathcal{S} \Big\{
\big[\bar s_a \gamma_{\alpha_1} \gamma_5 {\overset{\leftrightarrow}{D}}_{\alpha_3} s_a \big] {\overset{\leftrightarrow}{D}}_{\alpha_4} (\bar{s}_b \gamma_{\alpha_2} \gamma_5 s_b)
\Big\} ,
\\ \label{def:xi42}
\\[2mm] \nonumber \xi_{4^{+-};3}^{\alpha_1\alpha_2\alpha_3\alpha_4} &=&
g^{\mu\nu} \mathcal{S} \Big\{
\big[\bar s_a \sigma_{\alpha_1\mu} {\overset{\leftrightarrow}{D}}_{\alpha_3} s_a \big] {\overset{\leftrightarrow}{D}}_{\alpha_4} (\bar{s}_b \sigma_{\alpha_2\nu} s_b)
\Big\} .
\\ \label{def:xi43}
\end{eqnarray}
We can similarly construct three independent meson-meson currents of $J^{PC} = 0^{+-}$:
\begin{eqnarray}
\xi_{0^{+-};1} &=& \mathcal{P}_{J=0}^{\alpha_1\alpha_2\alpha_3\alpha_4}
\label{def:xi01}
\\ \nonumber &\times&
\big[\bar s_a \gamma_{\alpha_1} {\overset{\leftrightarrow}{D}}_{\alpha_3} s_a \big] {\overset{\leftrightarrow}{D}}_{\alpha_4} (\bar{s}_b \gamma_{\alpha_2} s_b),
\\[2mm] \xi_{0^{+-};2} &=& \mathcal{P}_{J=0}^{\alpha_1\alpha_2\alpha_3\alpha_4}
\label{def:xi02}
\\ \nonumber &\times&
\big[\bar s_a \gamma_{\alpha_1} \gamma_5 {\overset{\leftrightarrow}{D}}_{\alpha_3} s_a \big] {\overset{\leftrightarrow}{D}}_{\alpha_4} (\bar{s}_b \gamma_{\alpha_2} \gamma_5 s_b),
\\[2mm] \xi_{0^{+-};3} &=& \mathcal{P}_{J=0}^{\alpha_1\alpha_2\alpha_3\alpha_4} \times g^{\mu\nu}
\label{def:xi03}
\\ \nonumber &\times&
\big[\bar s_a \sigma_{\alpha_1\mu} {\overset{\leftrightarrow}{D}}_{\alpha_3} s_a \big] {\overset{\leftrightarrow}{D}}_{\alpha_4} (\bar{s}_b \sigma_{\alpha_2\nu} s_b),
\end{eqnarray}
and three independent meson-meson currents of $J^{PC} = 2^{+-}$:
\begin{eqnarray}
\xi_{2^{+-};1}^{\beta_1 \beta_2} &=& \mathcal{P}_{J=2;\beta_1 \beta_2}^{\alpha_1\alpha_2\alpha_3\alpha_4}
\label{def:xi21}
\\ \nonumber &\times&
\big[\bar s_a \gamma_{\alpha_1} {\overset{\leftrightarrow}{D}}_{\alpha_3} s_a \big] {\overset{\leftrightarrow}{D}}_{\alpha_4} (\bar{s}_b \gamma_{\alpha_2} s_b),
\\[2mm] \xi_{2^{+-};2}^{\beta_1 \beta_2} &=& \mathcal{P}_{J=2;\beta_1 \beta_2}^{\alpha_1\alpha_2\alpha_3\alpha_4}
\label{def:xi22}
\\ \nonumber &\times&
\big[\bar s_a \gamma_{\alpha_1} \gamma_5 {\overset{\leftrightarrow}{D}}_{\alpha_3} s_a \big] {\overset{\leftrightarrow}{D}}_{\alpha_4} (\bar{s}_b \gamma_{\alpha_2} \gamma_5 s_b),
\\[2mm] \xi_{2^{+-};3}^{\beta_1 \beta_2} &=& \mathcal{P}_{J=2;\beta_1 \beta_2}^{\alpha_1\alpha_2\alpha_3\alpha_4} \times g^{\mu\nu}
\label{def:xi23}
\\ \nonumber &\times&
\big[\bar s_a \sigma_{\alpha_1\mu} {\overset{\leftrightarrow}{D}}_{\alpha_3} s_a \big] {\overset{\leftrightarrow}{D}}_{\alpha_4} (\bar{s}_b \sigma_{\alpha_2\nu} s_b).
\end{eqnarray}
As depicted in Fig.~\ref{fig:relation}, the internal orbital angular momenta contained in the above meson-meson currents are all
\begin{equation}
\xi_{0/2/4^{+-};1/2/3}^{\cdots} : L=2 ; \, l_\lambda^\prime = 1 , \, l_\rho^\prime = 1 , \, l_{\rho^\prime}^\prime = 0 ,
\end{equation}
where $l^\prime_\rho$ and $l^\prime_{\rho^\prime}$ are the momenta inside the two mesons, and $l^\prime_\lambda$ is the momentum between them.

We can apply the Fierz rearrangement to derive the relations between $\eta^{\cdots}_{0/2/4^{+-};1/2/3}$ and $\xi^{\cdots}_{0/2/4^{+-};1/2/3}$ to be
\begin{equation}
\left(\begin{array}{c}
\eta^{\cdots}_{0/2/4^{+-};1}
\\
\eta^{\cdots}_{0/2/4^{+-};2}
\\
\eta^{\cdots}_{0/2/4^{+-};3}
\end{array}\right)
=
\left(\begin{array}{ccc}
2 & -2 & -2
\\
-2 & 2 & -2
\\
-4 & -4 & 0
\end{array}\right)
\left(\begin{array}{c}
\xi^{\cdots}_{0/2/4^{+-};1}
\\
\xi^{\cdots}_{0/2/4^{+-};2}
\\
\xi^{\cdots}_{0/2/4^{+-};3}
\end{array}\right)
\, .
\label{eq:fierz}
\end{equation}
We shall use these Fierz identities to study the decay behaviors at the end of this paper.

\section{QCD sum rule analyses}
\label{sec:sumrule}

The QCD sum rule method has been successfully applied to study various conventional and exotic hadrons in the past fifty years~\cite{Shifman:1978bx,Reinders:1984sr,Narison:2002woh,Nielsen:2009uh,Chen:2015moa,Chen:2020aos,Chen:2020uif,Chen:2021erj}. In this section we apply this non-perturbative method to study the fully-strange tetraquark currents $\eta_{0^{+-};1/2/3}$ of $J^{PC} = 0^{+-}$ and $\eta_{2^{+-};1/2/3}^{\beta_1 \beta_2}$ of $J^{PC} = 2^{+-}$.

We use the three currents $\eta_{0^{+-};1/2/3}$ of $J^{PC} = 0^{+-}$ as examples, and assume that they couple to the states $X_n$ ($n=1 \cdots N$) through
\begin{equation}
\langle 0| \eta_{0^{+-};i} | X_n \rangle = f_{in} \, ,
\label{eq:defg}
\end{equation}
where $X_n$ is the state that the current $\eta_{0^{+-};i}$ can couple to, $N$ is the number of such states, and $f_{in}$ is the $3 \times N$ matrix for the coupling of the current $\eta_{0^{+-};i}$ to the state $X_n$. Then we can investigate the diagonal and off-diagonal correlation functions
\begin{equation}
\Pi_{ij}(q^2) = i \int d^4x e^{iqx} \langle 0 | {\bf T}[ \eta_{0^{+-};i}(x) { \eta_{0^{+-};j}^\dagger } (0)] | 0 \rangle \, ,
\label{def:pi}
\end{equation}
at both the hadron and quark-gluon levels.

At the hadron level we use the dispersion relation to express $\Pi_{ij}(q^2)$ as
%
\begin{equation}
\Pi_{ij}(q^2) = \int^\infty_{s_<}\frac{\rho^{\rm phen}_{ij}(s)}{s-q^2-i\varepsilon}ds \, ,
\end{equation}
%
where $s_< = 16 m_s^2$ is the physical threshold. The spectral density $\rho^{\rm phen}_{ij}(s)$ can be generally parameterized for the states $X_n$ and a continuum as
%
\begin{eqnarray}
\nonumber&&\rho^{\rm phen}_{ij}(s)
\\ \nonumber
&=& \sum_n \delta(s-M^2_n) \langle 0| \eta_{0^{+-};i} | X_n \rangle \langle X_n | \eta_{0^{+-};j}^\dagger |0 \rangle + \cdots
\\ &=& \sum_n f_{in}f_{jn} \delta(s-M^2_n) + \cdots \, ,
\label{eq:rho}
\end{eqnarray}
where $M_n$ is the mass of the state $X_n$.

\begin{figure}[hbtp]
\begin{center}
\scalebox{0.12}{\includegraphics{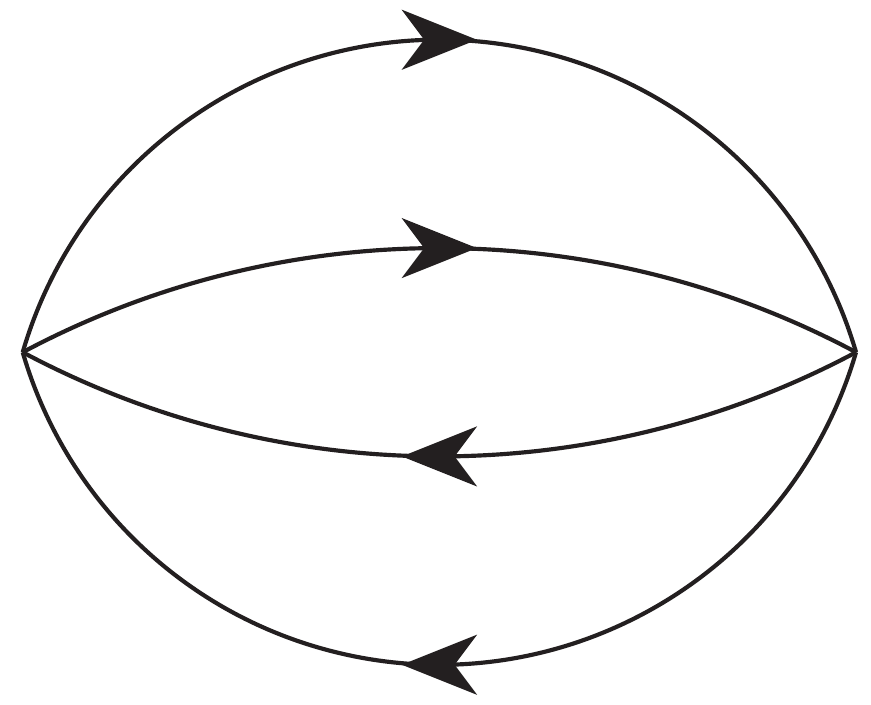}}
\\[2mm]
\scalebox{0.12}{\includegraphics{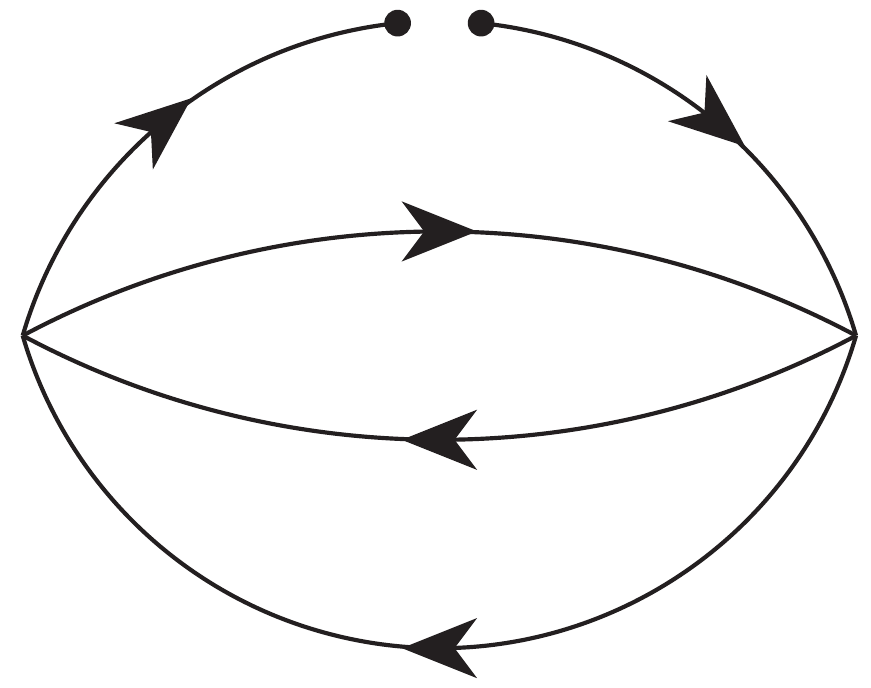}}~~
\scalebox{0.12}{\includegraphics{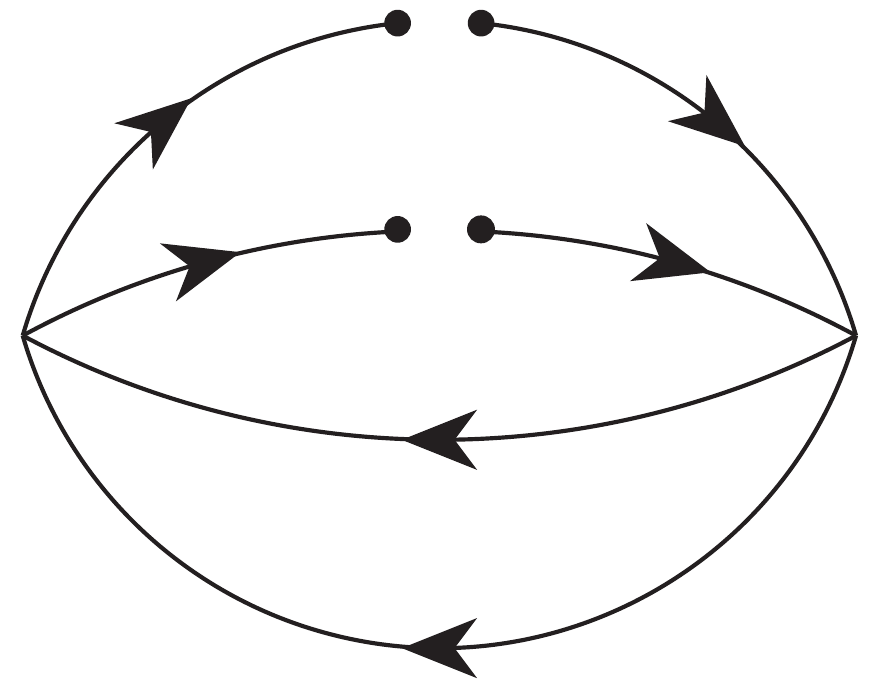}}~~
\scalebox{0.12}{\includegraphics{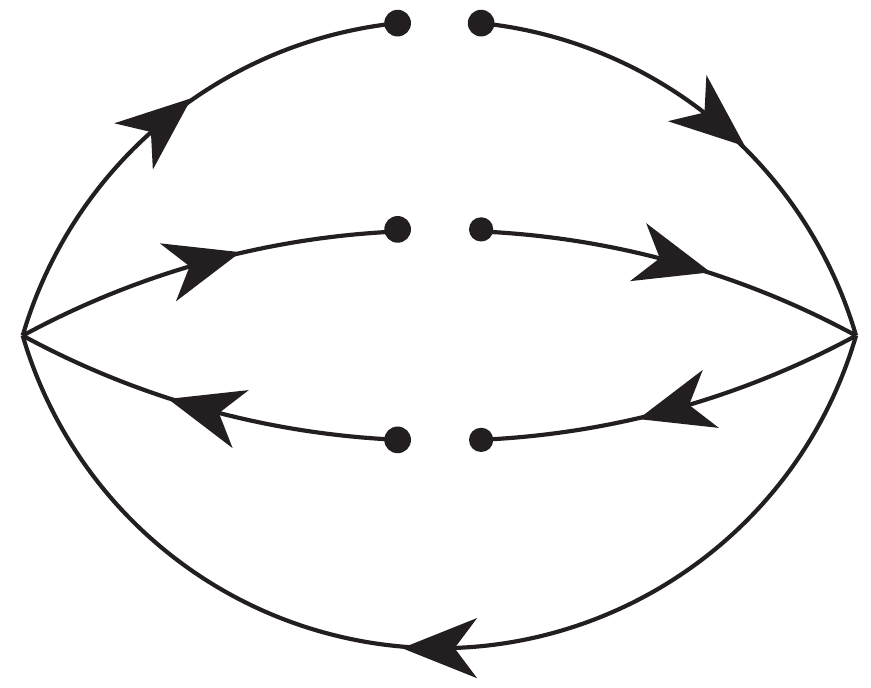}}
\\[2mm]
\scalebox{0.12}{\includegraphics{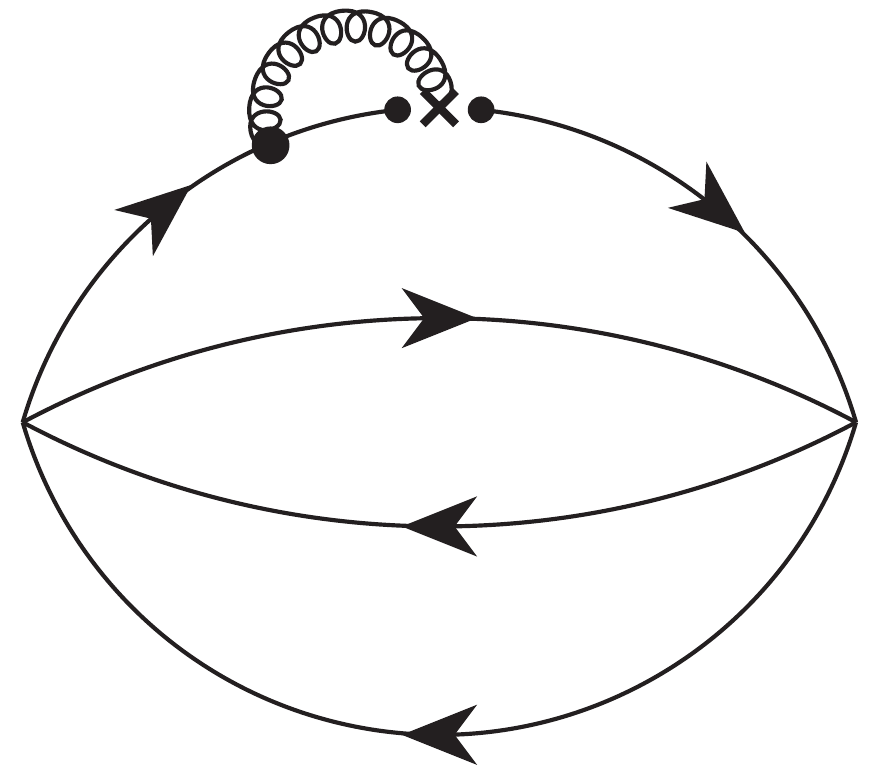}}~~
\scalebox{0.12}{\includegraphics{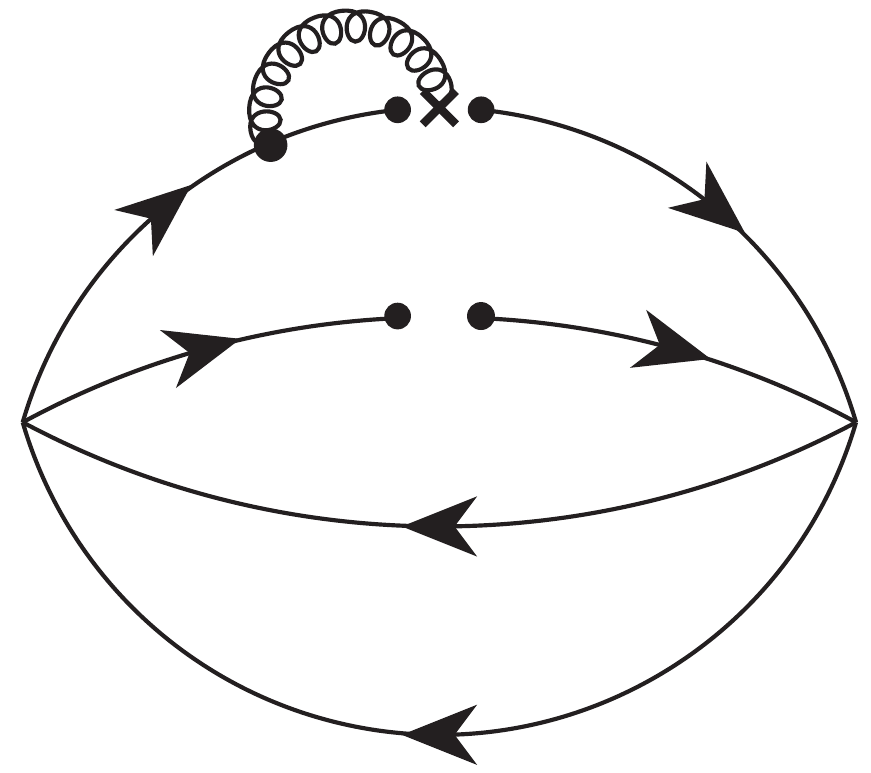}}~~
\scalebox{0.12}{\includegraphics{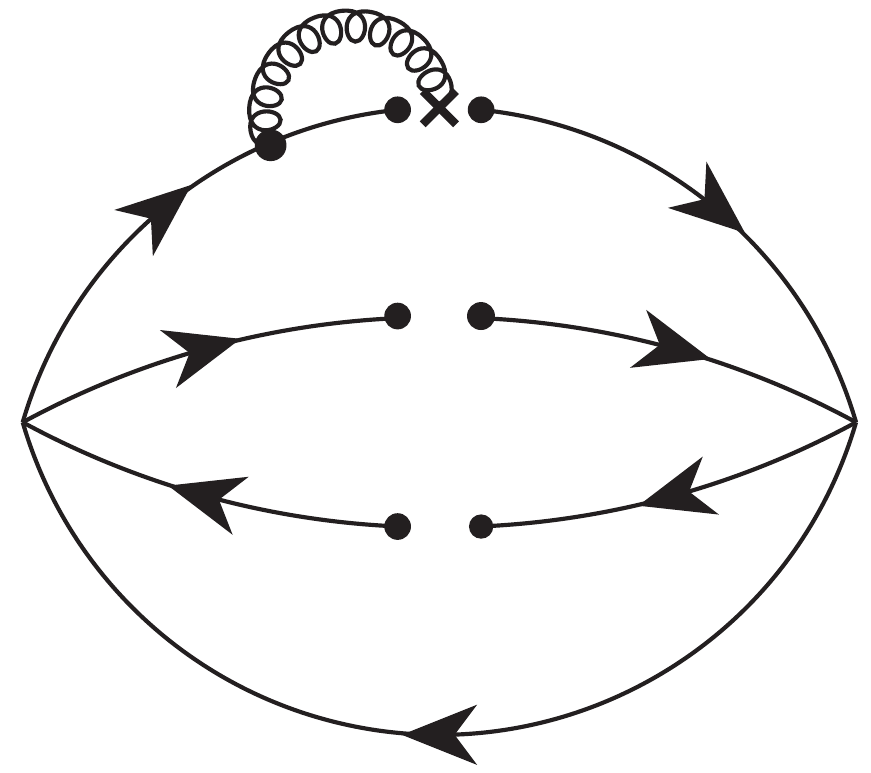}}~~
\scalebox{0.12}{\includegraphics{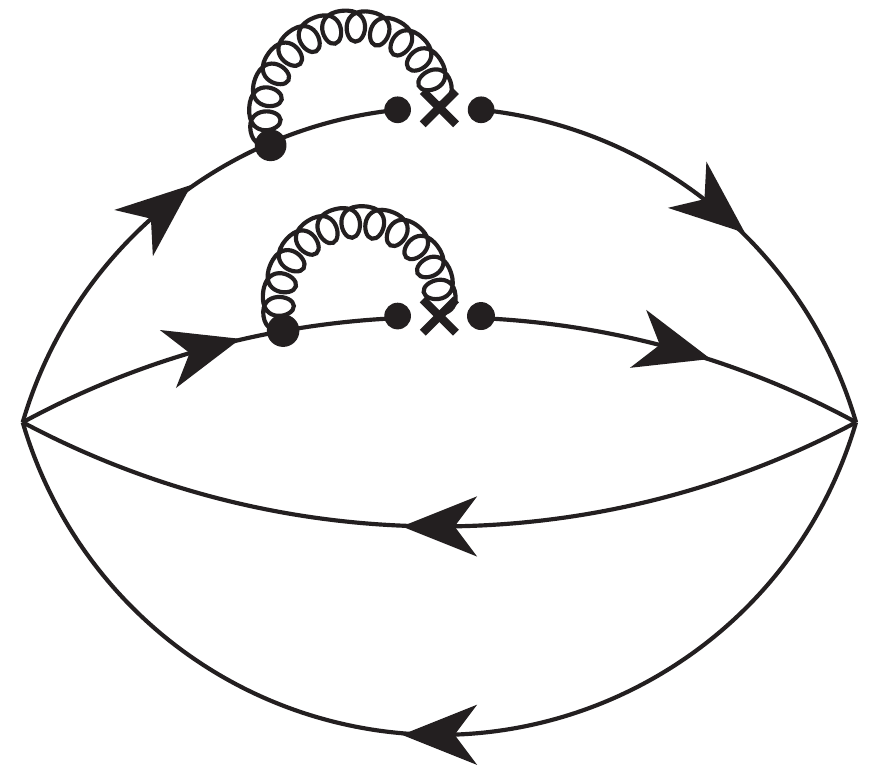}}
\\[2mm]
\scalebox{0.12}{\includegraphics{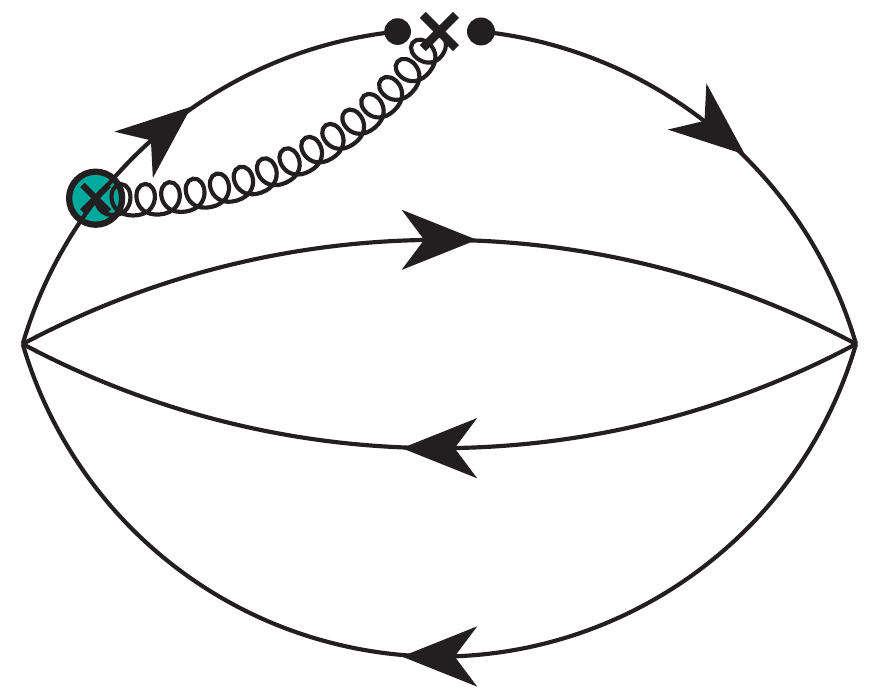}}~~
\scalebox{0.12}{\includegraphics{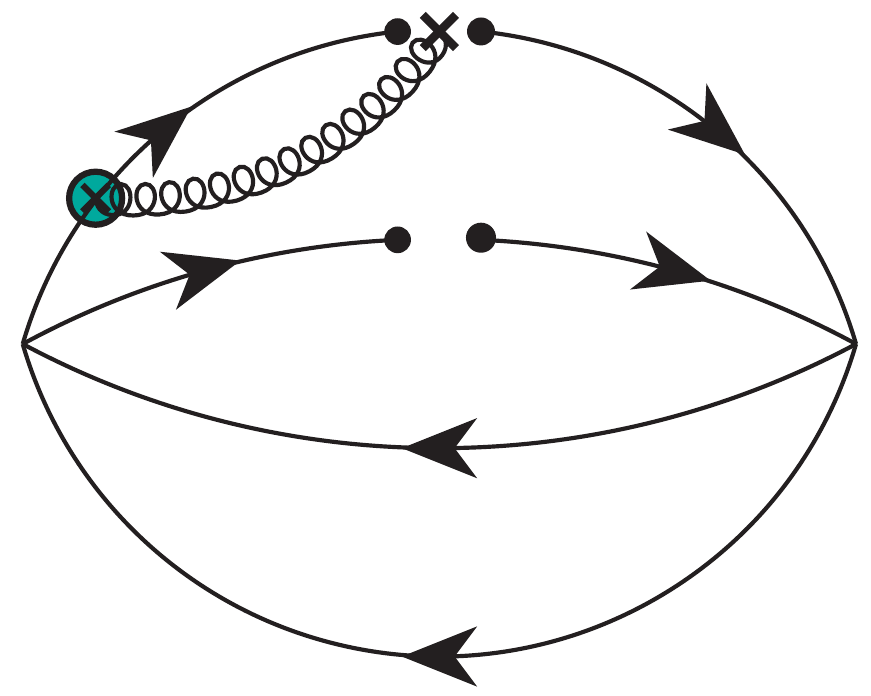}}~~
\scalebox{0.12}{\includegraphics{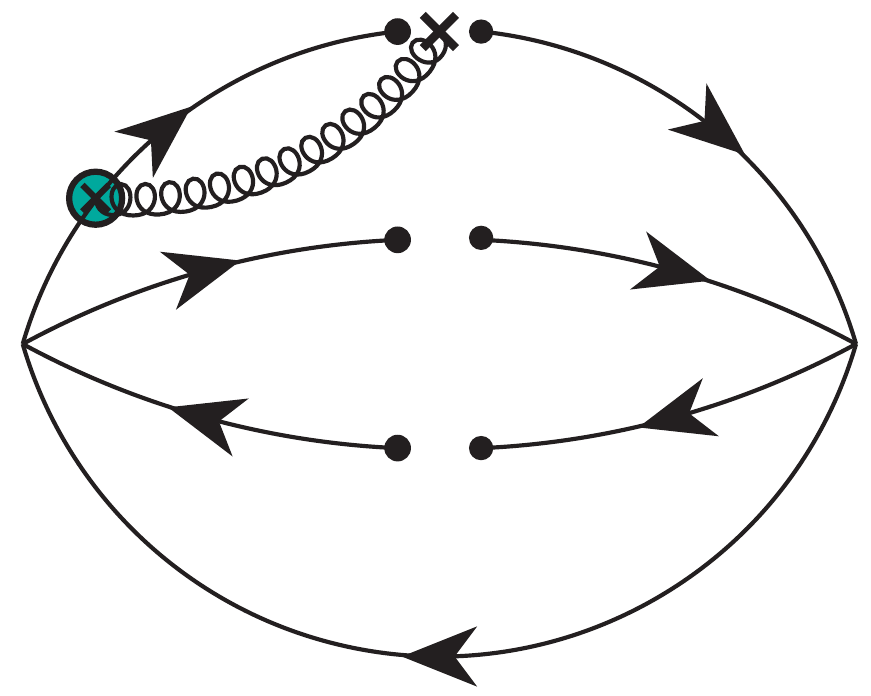}}~~
\scalebox{0.12}{\includegraphics{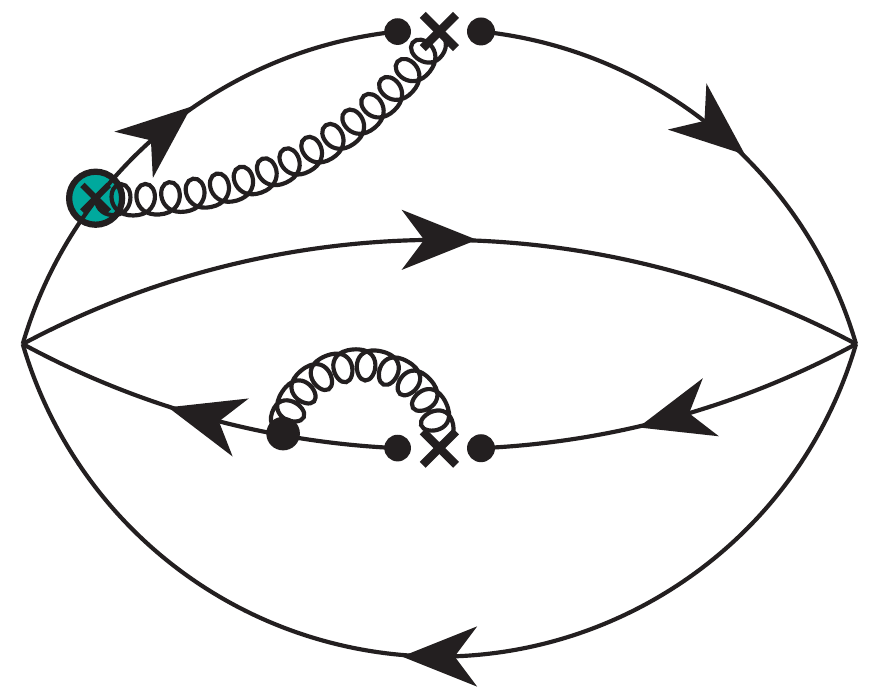}}
\\[2mm]
\scalebox{0.12}{\includegraphics{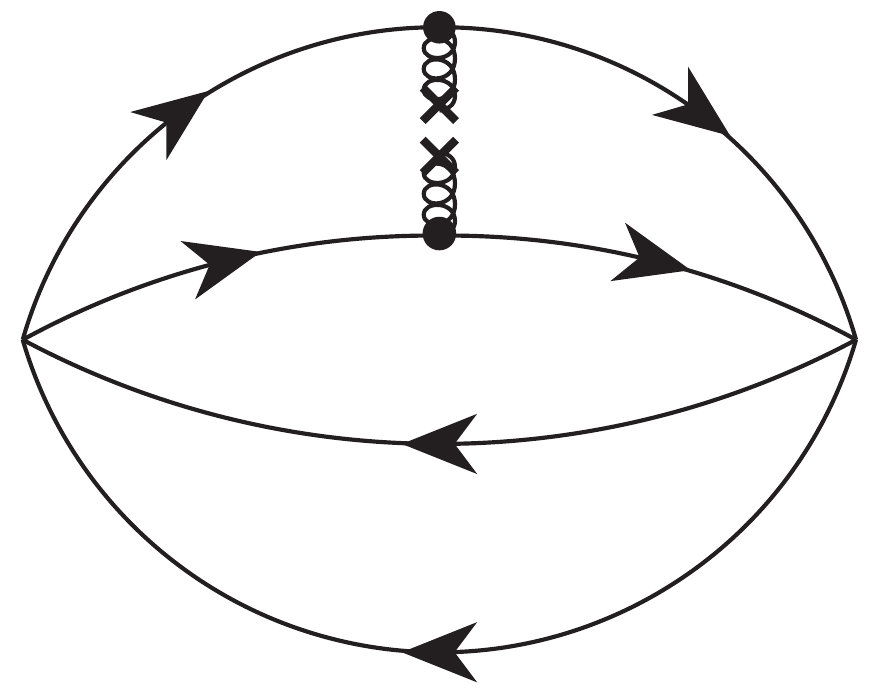}}~~
\scalebox{0.12}{\includegraphics{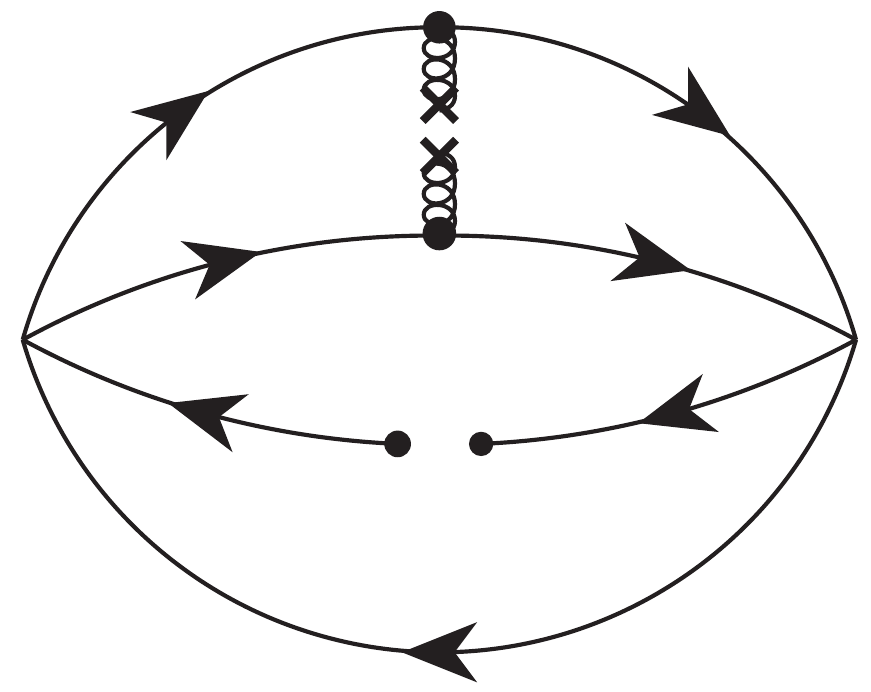}}~~
\scalebox{0.12}{\includegraphics{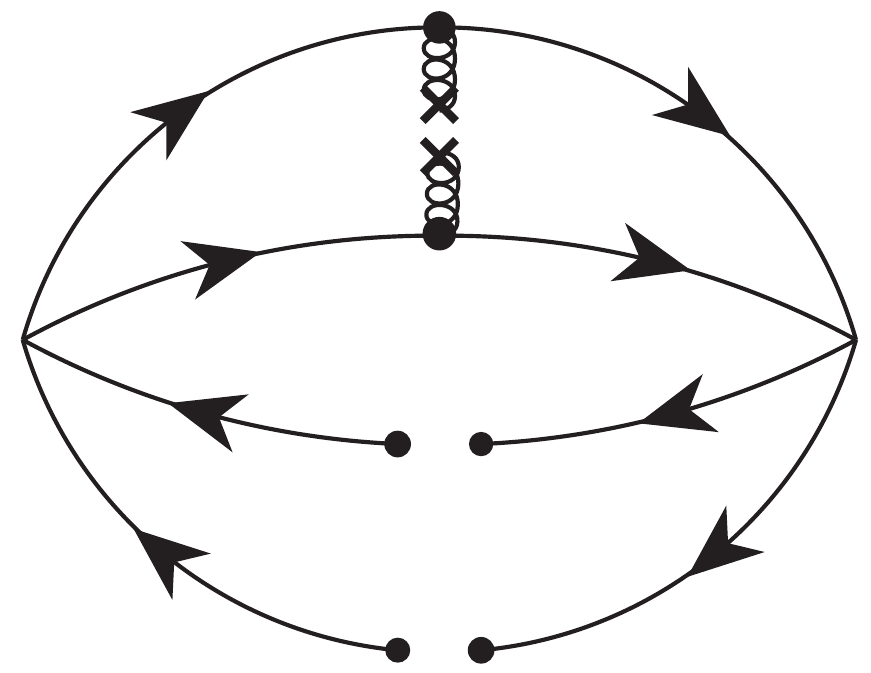}}
\\[2mm]
\scalebox{0.12}{\includegraphics{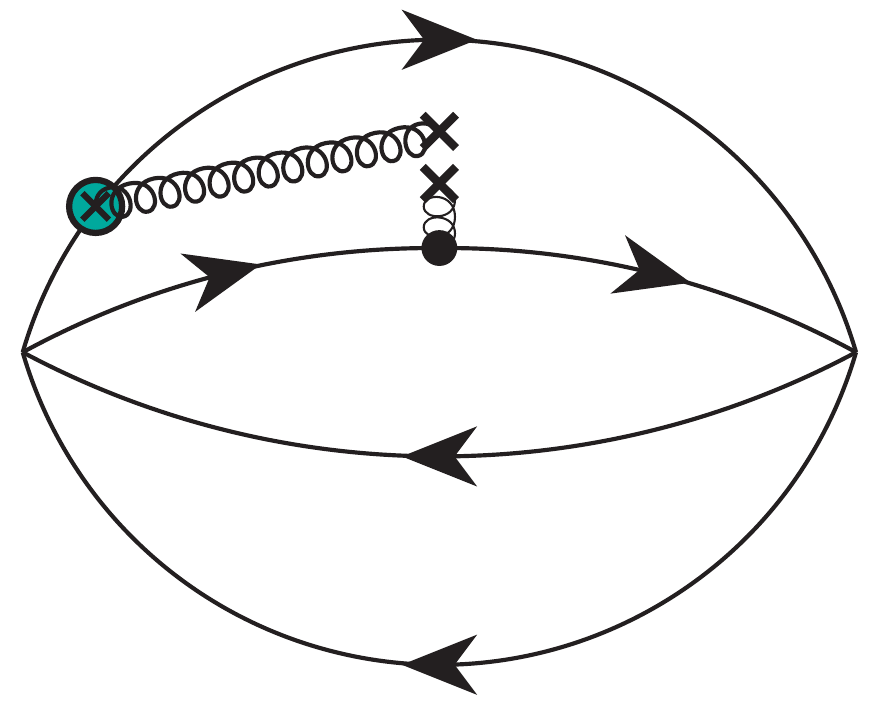}}~~
\scalebox{0.12}{\includegraphics{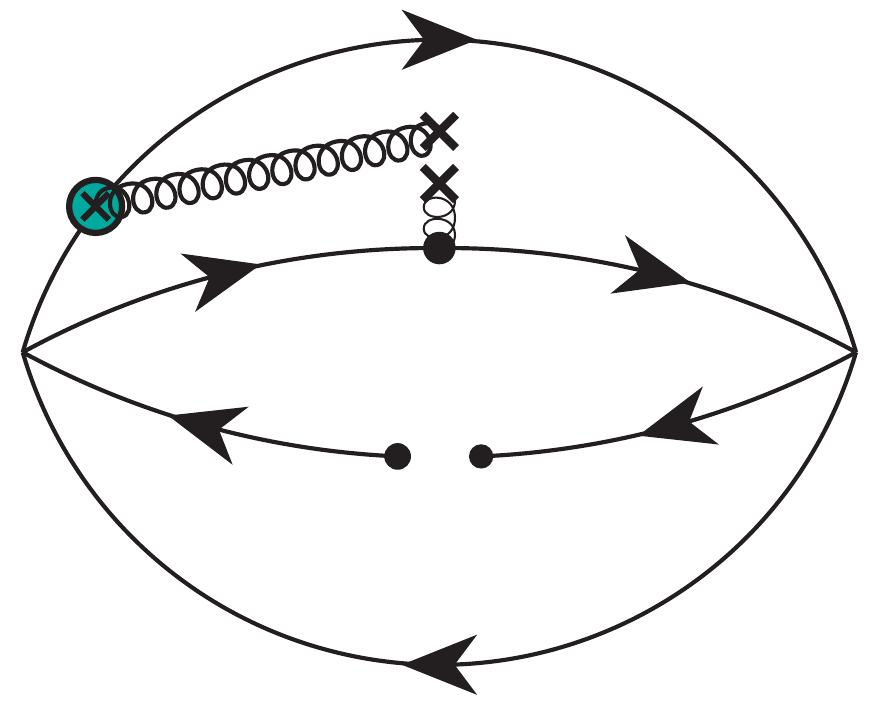}}~~
\scalebox{0.12}{\includegraphics{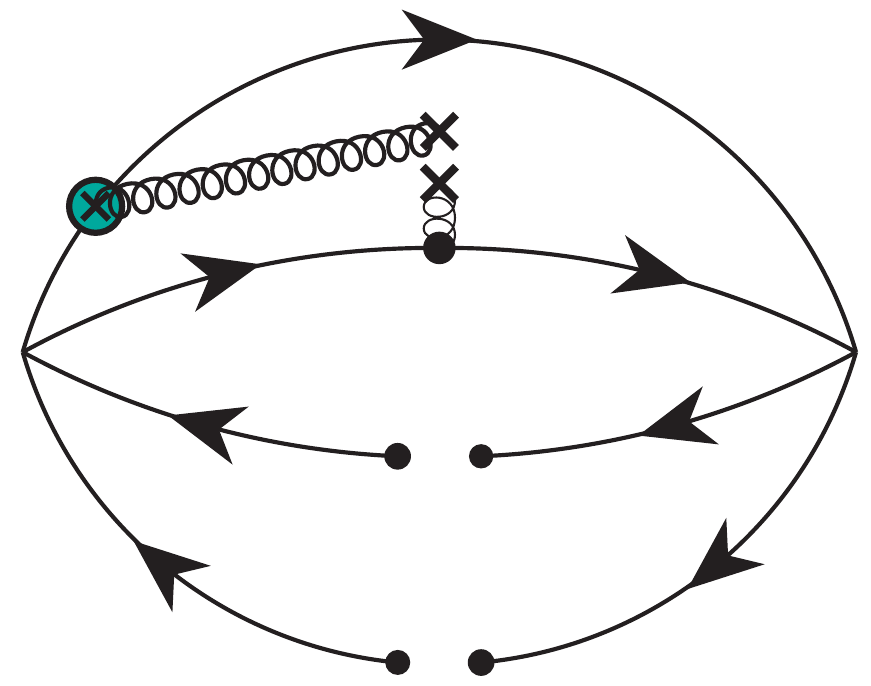}}
\\[2mm]
\scalebox{0.12}{\includegraphics{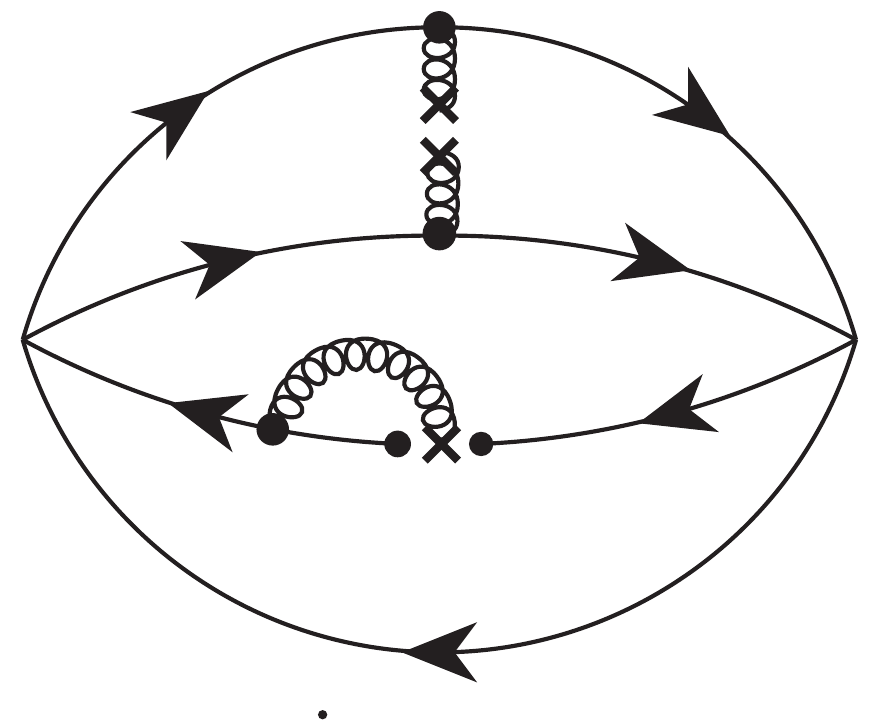}}~~
\scalebox{0.12}{\includegraphics{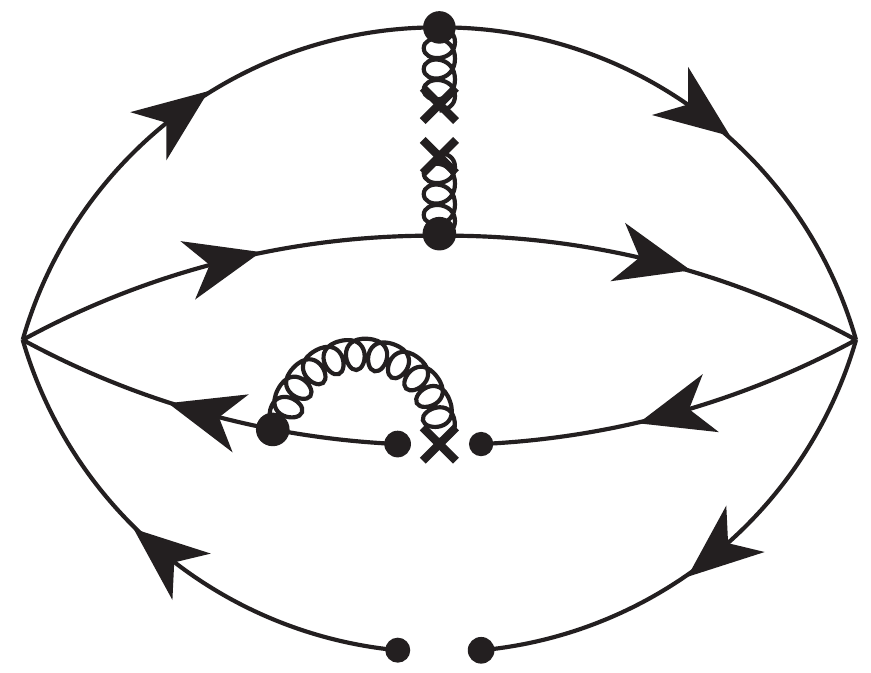}}~~
\scalebox{0.12}{\includegraphics{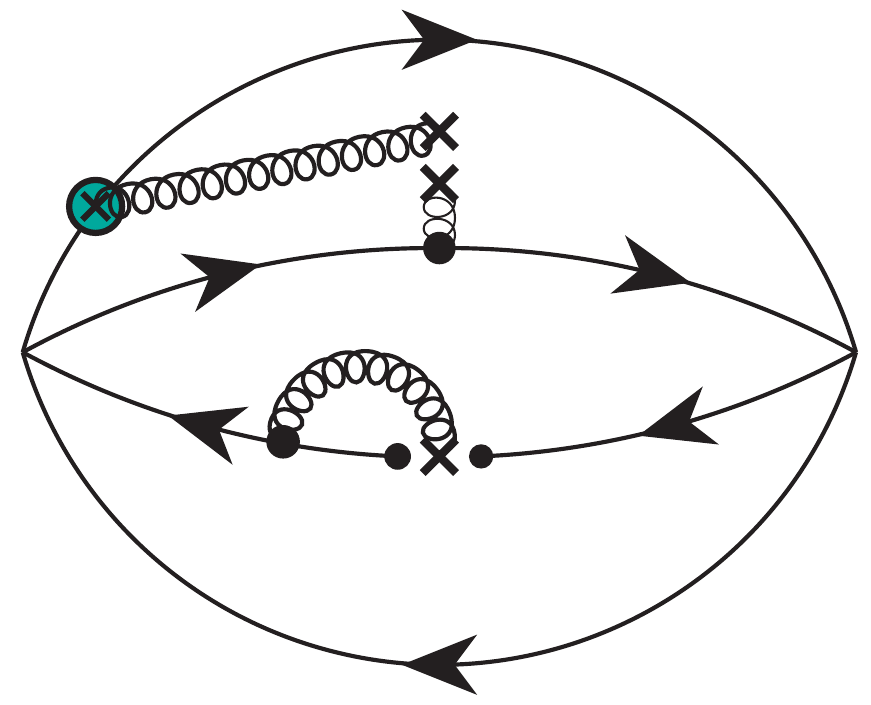}}~~
\scalebox{0.12}{\includegraphics{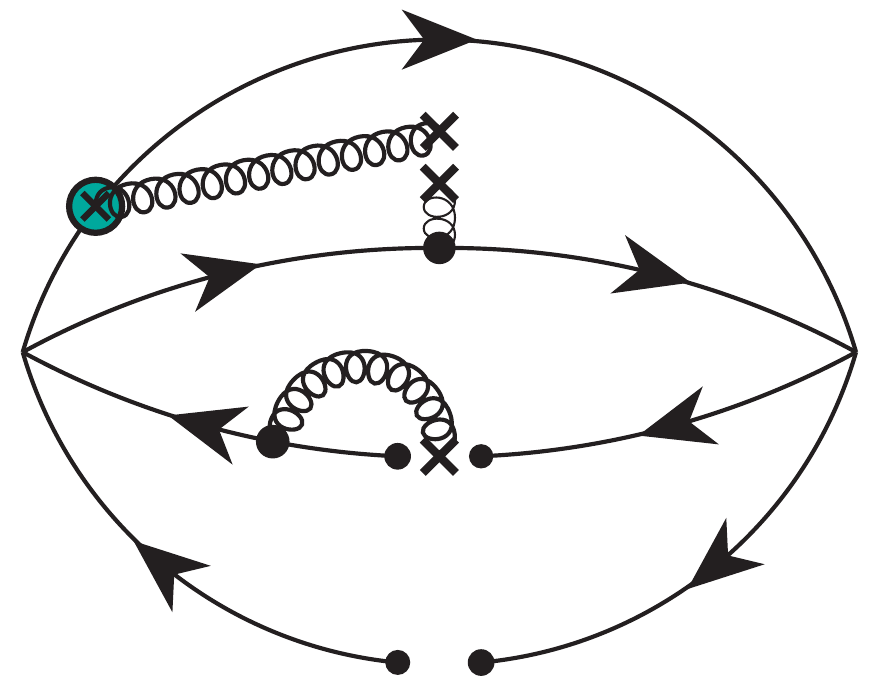}}
\caption{Feynman diagrams calculated in the present study. The covariant derivative $D_\alpha = \partial_\alpha + i g_s A_\alpha$ contains two terms, and we use the green vertex to describe the latter term.}
\label{fig:feynman}
\end{center}
\end{figure}

At the quark-gluon level we apply the method of operator product expansion (OPE) to calculate $\Pi_{ij}(q^2)$, from which we can extract the OPE spectral density $\rho_{ij}(s) \equiv \rho^{\rm OPE}_{ij}(s)$. In the present study we have calculated the Feynman diagrams depicted in Fig.~\ref{fig:feynman}, where we use the strangeness quark propagator as
\begin{eqnarray}
&&iS^{ab}_s(x-y)
\\ \nonumber
&=& \langle 0 | {\bf T}[ s^a(x) \bar s^b (y)] | 0 \rangle
\\ \nonumber &=& \frac {i \delta^{ab}}{ 2 \pi^2 (x-y)^4} (\hat{x}-\hat{y})
- \frac {\delta^{ab}}{12} \langle \bar s s \rangle
\\ \nonumber &&
+ \frac {i}{ 32 \pi^2 (x-y)^2}  \frac{\lambda^{n}_{ab}}{2} g_s G^{n}_{\mu \nu}  \left(\sigma^{\mu \nu} (\hat{x}-\hat{y}) + (\hat{x}-\hat{y}) \sigma^{\mu \nu} \right)
\\ \nonumber &&
-\frac {1}{ 4 \pi^2 (x-y)^4}  \frac{\lambda^{n}_{ab}}{2} g_s G^{n}_{\mu \nu} x^\mu y^\nu (\hat{x}-\hat{y})
\\ \nonumber &&
+  \frac { \delta^{ab} (x-y)^2}{192} \langle g_s \bar s \sigma G s \rangle
- \frac {m_s \delta^{ab}}{4 \pi^2 (x-y)^2}
\\ \nonumber &&
+ \frac {i \delta^{ab}}{48} m_s \langle \bar s s \rangle (\hat{x}-\hat{y}) + \frac {i \delta^{ab}}{8 \pi^2 (x-y)^2} m^{2}_s (\hat{x}-\hat{y})
\, .
\end{eqnarray}
We have considered the perturbative term, the quark condensate $\langle \bar s s \rangle$, the gluon condensate $\langle g_s^2 GG \rangle$, the quark-gluon mixed condensate $\langle g_s \bar s \sigma G s \rangle$, and their combinations. We have calculated all the diagrams proportional to $g_s^{N=0}$ and $g_s^{N=1}$, but we have only partly calculated the diagrams proportional to $g_s^{N\geq2}$.

\begin{widetext}
The OPE spectral densities $\rho_{ij}(s) \equiv \rho_{0^{+-};ij}(s)$ extracted from the three $J^{PC} = 0^{+-}$ currents $\eta_{0^{+-};1/2/3}$ are
\begin{eqnarray}
\label{eq:rhoeta1}
\rho_{0^{+-};11}(s)&=& {s^6 \over 1433600 \pi^6}+{7 m_s^2 s^5 \over 76800 \pi^6}
-{3 m_s \langle \bar s s \rangle \over 640 \pi^4} s^4
\\ \nonumber &&
- \Big (-{7 \langle g_s^2 GG \rangle m_s^2 \over 12288 \pi^6}-{{\langle \bar s s \rangle}^2 \over 24 \pi^2}+{37 m_s \langle g_s \bar s \sigma G s \rangle \over 2304 \pi^4} \Big ) s^3
\\ \nonumber &&
-\Big ({43 \langle g_s^2 GG \rangle m_s \langle \bar s s \rangle \over 2304 \pi^4}+{83 m_s^2 {\langle \bar s s \rangle}^2 \over 48 \pi^2}
-{11 \langle \bar s s \rangle \langle g_s \bar s \sigma G s \rangle \over 48 \pi^2} \Big ) s^2
\\ \nonumber &&
-\Big (-{\langle g_s^2 GG \rangle {\langle \bar s s \rangle}^2 \over 32 \pi^2} +{23  \langle g_s^2 GG \rangle m_s \langle g_s \bar s \sigma G s \rangle \over 768 \pi^4}+{97 m_s^2 \langle \bar s s \rangle \langle g_s \bar s \sigma G s \rangle \over 24 \pi^2}-{{\langle g_s \bar s \sigma G s \rangle}^2 \over 8 \pi^2} \Big ) s
\\ \nonumber &&
+{\langle g_s^2 GG \rangle m_s^2 {\langle \bar s s \rangle}^2 \over 72 \pi^2}+{\langle g_s^2 GG \rangle \langle \bar s s \rangle \langle g_s \bar s \sigma G s \rangle \over 32 \pi^2}-{25 m_s^2 {\langle g_s \bar s \sigma G s \rangle}^2 \over 24 \pi^2}
\,,
\\ \label{eq:rhoeta4}
\rho_{0^{+-};22}(s) &=& {s^6 \over 716800 \pi^6}-{m_s^2 s^5 \over 3840 \pi^6}-\Big (-{\langle g_s^2 GG \rangle \over 30720 \pi^6}-{11 m_s \langle \bar s s \rangle \over 960 \pi^4} \Big ) s^4
\\ \nonumber &&
-\Big (-{3 \langle g_s^2 GG \rangle m_s^2 \over 4096 \pi^6}-{77 m_s \langle g_s \bar s \sigma G s \rangle \over 2304 \pi^4}+{{\langle \bar s s \rangle}^2 \over 12 \pi^2}\Big ) s^3
\\ \nonumber &&
-\Big ({49 \langle g_s^2 GG \rangle m_s \langle \bar s s \rangle \over 2304 \pi^4}-{29 m_s^2 {\langle \bar s s \rangle}^2 \over 24 \pi^2}
+{19 \langle \bar s s \rangle \langle g_s \bar s \sigma G s \rangle \over 48 \pi^2} \Big ) s^2
\\ \nonumber &&
-\Big ({23 \langle g_s^2 GG \rangle m_s \langle g_s \bar s \sigma G s \rangle \over 768 \pi^4}-{ \langle g_s^2 GG \rangle {\langle \bar s s \rangle}^2 \over 32 \pi^2}+{43 m_s^2 \langle \bar s s \rangle \langle g_s \bar s \sigma G s \rangle \over 24 \pi^2}+{5 {\langle g_s \bar s \sigma G s \rangle}^2 \over 32 \pi^2} \Big ) s
\\ \nonumber &&
+{\langle g_s^2 GG \rangle \langle \bar s s \rangle \langle g_s \bar s \sigma G s \rangle \over 32 \pi^2}-{43 m_s^2 {\langle g_s \bar s \sigma G s \rangle}^2 \over 24 \pi^2}+{7  \langle g_s^2 GG \rangle m_s^2 {\langle \bar s s \rangle}^2 \over 144 \pi^2}
\,,
\\ \label{eq:rhoeta5}
\rho_{0^{+-};33}(s) &=& {s^6 \over 179200 \pi^6}-{m_s^2 s^5 \over 1600 \pi^6}-\Big (-{\langle g_s^2 GG \rangle \over 12288 \pi^6}-{m_s \langle \bar s s \rangle \over 96 \pi^4} \Big ) s^4- {5 \langle g_s^2 GG \rangle m_s^2 \over 1536 \pi^6} s^3
\\ \nonumber &&
+{\langle g_s^2 GG \rangle m_s \langle \bar s s \rangle \over 36 \pi^4} s^2 - {5 m_s^2 \langle \bar s s \rangle \langle g_s \bar s \sigma G s \rangle \over \pi^2} s
\\ \nonumber &&
-{2 m_s^2 {\langle g_s \bar s \sigma G s \rangle}^2 \over \pi^2}+{31  \langle g_s^2 GG \rangle m_s^2 {\langle \bar s s \rangle}^2 \over 144 \pi^2}
\,,
\\ \label{eq:rhoeta14}
\rho_{0^{+-};12}(s) &=& {\langle g_s^2 GG \rangle s^4 \over 122880 \pi^6}-\Big ({7 \langle g_s^2 GG \rangle m_s^2 \over 6144 \pi^6}-{5 m_s \langle g_s \bar s \sigma G s \rangle \over 256 \pi^4} \Big ) s^3
\\ \nonumber &&
-\Big (-{\langle g_s^2 GG \rangle m_s \langle \bar s s \rangle \over 768 \pi^4}+{\langle \bar s s \rangle \langle g_s \bar s \sigma G s \rangle \over 4 \pi^2} \Big ) s^2
\\ \nonumber &&
-\Big (-{9 m_s^2 \langle \bar s s \rangle \langle g_s \bar s \sigma G s \rangle \over 8 \pi^2}+{3 {\langle g_s \bar s \sigma G s \rangle}^2 \over 8 \pi^2} \Big ) s
\\ \nonumber &&
+{\langle g_s^2 GG \rangle m_s^2 {\langle \bar s s \rangle}^2 \over 32 \pi^2}+{15 m_s^2 {\langle g_s \bar s \sigma G s \rangle}^2 \over 32 \pi^2}
\,,
\\ \label{eq:rhoeta15}
\rho_{0^{+-};13}(s) &=& -{41 m_s^2 s^5 \over 153600 \pi^6} + {m_s \langle \bar s s \rangle \over 96 \pi^4} s^4 - \Big ({5 \langle g_s^2 GG \rangle m_s^2 \over 9216 \pi^6} -{545 m_s \langle g_s \bar s \sigma G s \rangle \over 9216 \pi^4} \Big ) s^3
\\ \nonumber &&
-\Big (-{\langle g_s^2 GG \rangle m_s \langle \bar s s \rangle \over 192 \pi^4}-{5 m_s^2 {\langle \bar s s \rangle}^2 \over 3 \pi^2} \Big )  s^2
\\ \nonumber &&
-\Big (-{\langle g_s^2 GG \rangle m_s \langle g_s \bar s \sigma G s \rangle \over 128 \pi^4}-{\langle g_s^2 GG \rangle {\langle \bar s s \rangle}^2 \over 192 \pi^2}-{523 m_s^2 \langle \bar s s \rangle \langle g_s \bar s \sigma G s \rangle \over 192 \pi^2} +{{\langle g_s \bar s \sigma G s \rangle}^2 \over 384 \pi^2} \Big ) s
\\ \nonumber &&
+{\langle g_s^2 GG \rangle \langle \bar s s \rangle \langle g_s \bar s \sigma G s \rangle \over 144 \pi^2}+{m_s^2 {\langle g_s \bar s \sigma G s \rangle}^2 \over 48 \pi^2}
\,,
\\ \label{eq:rhoeta45}
\rho_{0^{+-};23}(s) &=& \Big ({5 \langle g_s^2 GG \rangle m_s^2 \over 3072 \pi^6}-{37 m_s \langle g_s \bar s \sigma G s \rangle \over 3072 \pi^4} \Big ) s^3 - {\langle g_s^2 GG \rangle m_s \langle \bar s s \rangle \over 64 \pi^4} s^2
\\ \nonumber &&
-\Big ({3 \langle g_s^2 GG \rangle m_s \langle g_s \bar s \sigma G s \rangle \over 128 \pi^4}+{\langle g_s^2 GG \rangle {\langle \bar s s \rangle}^2 \over 64 \pi^2}+{47 m_s^2 \langle \bar s s \rangle \langle g_s \bar s \sigma G s \rangle \over 64 \pi^2}-{9 {\langle g_s \bar s \sigma G s \rangle}^2 \over 128 \pi^2} \Big )s
\\ \nonumber &&
-{\langle g_s^2 GG \rangle \langle \bar s s \rangle \langle g_s \bar s \sigma G s \rangle \over 48 \pi^2} -{m_s^2 {\langle g_s \bar s \sigma G s \rangle}^2 \over 8 \pi^2}
\, ,
\end{eqnarray}
and the OPE spectral densities $\rho_{2^{+-};ij}(s)$ extracted from the three $J^{PC} = 2^{+-}$ currents $\eta_{2^{+-};1/2/3}^{\beta_1 \beta_2}$ are
\begin{eqnarray}
\label{eq:rhoeta21}
\rho_{2^{+-};11}(s) &=& {341 s^6 \over 43545600 \pi^6}-{31 m_s^2 s^5 \over 44800 \pi^6}
-\Big ({41 \langle g_s^2 GG \rangle \over 1161216\pi^6}+ {31 m_s \langle \bar s s \rangle \over 15120 \pi^4 } \Big ) s^4
\\ \nonumber &&
-\Big (-{29 \langle g_s^2 GG \rangle m_s^2 \over 69120 \pi^6}-{2 {\langle \bar s s \rangle}^2 \over 9 \pi^2} + {2537 m_s \langle g_s \bar s \sigma G s \rangle \over 25920 \pi^4} \Big ) s^3
\\ \nonumber &&
-\Big ( {7 \langle g_s^2 GG \rangle m_s \langle \bar s s \rangle \over 4320 \pi^4}+ {5 m_s^2 {\langle \bar s s \rangle}^2 \over 18 \pi^2} - {629 \langle \bar s s \rangle \langle g_s \bar s \sigma G s \rangle \over 540 \pi^2} \Big ) s^2
\\ \nonumber &&
-\Big ( {\langle g_s^2 GG \rangle {\langle \bar s s \rangle}^2 \over 36 \pi^2} + {17 \langle g_s^2 GG \rangle m_s \langle g_s \bar s \sigma G s \rangle \over 576 \pi^4}+ {155 m_s^2 \langle \bar s s \rangle \langle g_s \bar s \sigma G s \rangle \over 18 \pi^2} - {221 {\langle g_s \bar s \sigma G s \rangle}^2 \over 216 \pi^2} \Big ) s
\\ \nonumber &&
-{5 \langle g_s^2 GG \rangle m_s^2 {\langle \bar s s \rangle}^2 \over 81 \pi^2} - {\langle g_s^2 GG \rangle \langle \bar s s \rangle \langle g_s \bar s \sigma G s \rangle \over 81 \pi^2} - {116 m_s^2 {\langle g_s \bar s \sigma G s \rangle}^2 \over 27 \pi^2}
\,,
\\ \label{eq:rhoeta24}
\rho_{2^{+-};22}(s) &=& {341 s^6 \over 21772800 \pi^6} - {551 m_s^2 s^5 \over 201600 \pi^6} - \Big (-{157 \langle g_s^2 GG \rangle \over 829440\pi^6}- {611 m_s \langle \bar s s \rangle \over 7560 \pi^4} \Big ) s^4
\\ \nonumber &&
- \Big ({241\langle g_s^2 GG \rangle m_s^2 \over 41472 \pi^6} + {4 {\langle \bar s s \rangle}^2 \over 9 \pi^2} - {4693 m_s \langle g_s \bar s \sigma G s \rangle \over 25920 \pi^4} \Big ) s^3
\\ \nonumber &&
 - \Big (- {\langle g_s^2 GG \rangle m_s \langle \bar s s \rangle \over 135 \pi^4}- {47 m_s^2 {\langle \bar s s \rangle}^2 \over 5 \pi^2}+ {215 \langle \bar s s \rangle \langle g_s\bar s \sigma G s \rangle \over 108 \pi^2} \Big ) s^2
\\ \nonumber &&
- \Big  ( {\langle g_s^2 GG \rangle {\langle \bar s s \rangle}^2 \over 36 \pi^2}+ {17 \langle g_s^2 GG \rangle m_s \langle g_s\bar s \sigma G s \rangle \over 576 \pi^4} + {m_s^2 \langle \bar s s \rangle \langle g_s \bar s \sigma G s \rangle \over 2 \pi^2} + {349 {\langle g_s \bar s \sigma G s \rangle}^2 \over 216 \pi^2} \Big ) s
\\ \nonumber &&
- {\langle g_s^2 GG \rangle m_s^2 {\langle \bar s s \rangle}^2 \over 81 \pi^2}- {\langle g_s^2 GG \rangle \langle \bar s s \rangle \langle g_s \bar s \sigma G s \rangle \over 81 \pi^2} + {28 m_s^2 {\langle g_s\bar s \sigma G s \rangle}^2 \over 27 \pi^2}
\,,
\\ \label{eq:rhoeta25}
\rho_{2^{+-};33}(s) &=&{13 s^6 \over 1036800 \pi^6} - {331 m_s^2  s^5 \over 201600 \pi^6}-\Big (-{13 \langle g_s^2 GG \rangle \over 161280 \pi^6}- {23 m_s \langle \bar s s \rangle \over 756 \pi^4} \Big ) s^4
\\ \nonumber &&
- {5 \langle g_s^2 GG \rangle m_s^2 \over 10368 \pi^6} s^3-\Big ( {2 \langle g_s^2 GG \rangle m_s \langle \bar s s \rangle \over 135 \pi^4} - {49 m_s^2 {\langle \bar s s \rangle}^2 \over 15 \pi^2} \Big ) s^2
\\ \nonumber &&
- {76 m_s^2 \langle \bar s s \rangle \langle g_s \bar s \sigma G s \rangle \over 9 \pi^2} s -{8 \langle g_s^2 GG \rangle m_s^2 {\langle \bar s s \rangle}^2 \over 81 \pi^2}- {32 m_s^2 {\langle g_s \bar s \sigma G s \rangle}^2 \over 9 \pi^2}
\,,
\\ \label{eq:rhoeta214}
\rho_{2^{+-};12}(s) &=& -{3\langle g_s^2 GG \rangle s^4 \over 71680 \pi^6}-\Big ({131\langle g_s^2 GG \rangle m_s^2 \over 34560 \pi^6} + {31 m_s \langle g_s \bar s \sigma G s \rangle \over 2160 \pi^4} \Big ) s^3
\\ \nonumber &&
-\Big (- {133 \langle g_s^2 GG  \rangle m_s \langle \bar s s \rangle \over 2880 \pi^4}- {2 \langle \bar s s \rangle \langle g_s \bar s \sigma G s \rangle \over 45 \pi^2} \Big ) s^2-\Big ( {2 m_s^2 \langle \bar s s \rangle \langle g_s \bar s \sigma G s \rangle \over 9 \pi^2}- {{\langle g_s \bar s \sigma G s \rangle}^2 \over 18 \pi^2} \Big ) s
\\ \nonumber &&
+{7\langle g_s^2 GG \rangle m_s^2 {\langle \bar s s \rangle}^2 \over 27 \pi^2}
\,,
\\ \label{eq:rhoeta215}
\rho_{2^{+-};13}(s) &=& -{41 m_s^2 s^5 \over 80640 \pi^6} + {13 m_s \langle \bar s s \rangle \over 840 \pi^4} s^4 -\Big ({59 \langle g_s^2 GG \rangle m_s^2 \over 207360 \pi^6}- {2 {\langle \bar s s \rangle}^2 \over 15 \pi^2} - {53 m_s \langle g_s \bar s \sigma G s \rangle \over 1296 \pi^4}  \Big ) s^3
\\ \nonumber &&
-\Big (-{\langle g_s^2 GG \rangle m_s \langle \bar s s \rangle \over 480 \pi^4}- {44 m_s^2 {\langle \bar s s \rangle}^2 \over 15 \pi^2} - {481 \langle \bar s s \rangle \langle g_s \bar s \sigma G s \rangle \over 270 \pi^2}  \Big ) s^2
\\ \nonumber &&
-\Big ({\langle g_s^2 GG \rangle {\langle \bar s s \rangle}^2 \over 216 \pi^2}- {\langle g_s^2 GG \rangle m_s \langle g_s \bar s \sigma G s \rangle \over 72 \pi^4} + {199 m_s^2 \langle \bar s s \rangle \langle g_s \bar s \sigma G s \rangle \over 27 \pi^2} - {503 {\langle g_s \bar s \sigma G s \rangle}^2 \over 216 \pi^2}  \Big ) s
\\ \nonumber &&
-{\langle g_s^2 GG \rangle\langle \bar s s \rangle \langle g_s \bar s \sigma G s \rangle \over 27 \pi^2}- {140 m_s^2 {\langle g_s \bar s \sigma G s \rangle}^2 \over 27 \pi^2}
\,,
\\ \label{eq:rhoeta245}
\rho_{2^{+-};23}(s) &=&-\Big (-{59 \langle g_s^2 GG \rangle m_s^2 \over  69120 \pi^6} - {7 m_s  \langle g_s \bar s \sigma G s \rangle \over  1440 \pi^4} \Big ) s^3 -\Big ({ \langle g_s^2 GG \rangle m_s \langle \bar s s \rangle \over  160 \pi^4} - { \langle \bar s s \rangle  \langle g_s \bar s \sigma G s \rangle \over  90 \pi^2} \Big ) s^2
\\ \nonumber &&
-\Big (-{ \langle g_s^2 GG \rangle {\langle \bar s s \rangle}^2 \over  72 \pi^2} + {\langle g_s^2 GG \rangle m_s  \langle g_s \bar s \sigma G s \rangle \over  24 \pi^4} - {5 m_s^2 \langle \bar s s \rangle  \langle g_s \bar s \sigma G s \rangle \over  9 \pi^2} + {{\langle g_s \bar s \sigma G s \rangle}^2 \over 24 \pi^2} \Big ) s
\\ \nonumber &&
+ { \langle g_s^2 GG \rangle \langle \bar s s \rangle  \langle g_s \bar s \sigma G s \rangle \over  9 \pi^2}
\, .
\end{eqnarray}

\end{widetext}
We take the following values for various QCD parameters contained in the above expressions~\cite{pdg,Yang:1993bp,Gimenez:2005nt,Jamin:2002ev,Ioffe:2002be,Ovchinnikov:1988gk,Ellis:1996xc,Narison:2011xe,Narison:2018dcr}:
\begin{eqnarray}
\nonumber m_{s}(2~{\rm GeV}) &=&93_{-~5}^{+11} {\rm~MeV} \,
\\ \nonumber
\langle\bar ss\rangle &=& -(0.8\pm 0.1)\times(0.240 \mbox{ GeV})^3\, ,
\\
\left\langle g_{s}^{2} G G\right\rangle &=&(0.48 \pm 0.14) {\rm~GeV}^{4} \, ,
\\ \nonumber
\left\langle g_{s} \bar{s} \sigma G s\right\rangle &=&-M_{0}^{2} \times\langle\bar{s} s\rangle \, ,
\\ \nonumber
M_{0}^{2} &=&(0.8 \pm 0.2) {\rm~GeV}^{2} \, .
\label{eq:condensates}
\end{eqnarray}

After performing the Borel transformation at both the hadron and quark-gluon levels, we approximate the continuum using $\rho_{ij}(s)$ above the threshold value $s_0$ to obtain
\begin{eqnarray}
\nonumber \Pi_{ij}(s_0, M_B^2) &=& \sum_n f_{in}f_{jn} e^{-M_n^2/M_B^2}
\\ &=& \int^{s_0}_{s_<} e^{-s/M_B^2} \rho_{ij}(s) ds \, .
\label{eq:fin}
\end{eqnarray}
We shall separately perform the single-channel and multi-channel analyses in the following two subsections.

\subsection{Single-channel analysis}
\label{sec:single}

In this subsection we perform the single-channel analysis by setting $\rho_{ij}(s)|_{i \neq j} = 0$. This assumption neglects the off-diagonal correlation functions to make the three currents $\eta_{0^{+-};1/2/3}$ ``non-correlated'', {\it i.e.}, any two of them can not mainly couple to the same state $X$, otherwise,
\begin{eqnarray}
\nonumber &&\rho_{ij}(s)
\\ \nonumber
&=& \sum_n \delta(s-M^2_n) \langle 0| \eta_{0^{+-};i} | X_n \rangle \langle X_n | \eta_{0^{+-};j}^\dagger |0 \rangle + \cdots
\\ \nonumber &\approx& \delta(s-M^2_{X}) \langle 0| \eta_{0^{+-};i} | X \rangle \langle X | \eta_{0^{+-};j}^\dagger |0 \rangle + \cdots
\\ &\neq& 0 \, .
\end{eqnarray}
Accordingly, we further assume that there are three states $X_{1,2,3}$ separately corresponding to the three currents $\eta_{0^{+-};1/2/3}$ through
\begin{equation}
\langle 0| \eta_{0^{+-};i} | X_i \rangle = f_{ii} \, .
\end{equation}
We parameterize the spectral density $\rho_{ii}(s)$ as one pole dominance for the single state $X_i$ between the physical threshold $s_<$ and the threshold value $s_0$ as well as a continuum contribution above $s_0$. This simplifies Eq.~(\ref{eq:fin}) to be
\begin{equation}
\Pi_{ii}(s_0, M_B^2) = f_{ii}^2 e^{-M_{i}^2/M_B^2} = \int^{s_0}_{s_<} e^{-s/M_B^2} \rho_{ii}(s) ds \, ,
\end{equation}
and the mass $M_{i}$ can be calculated through
%
\begin{equation}
M^2_{i}(s_0, M_B) = \frac{\int^{s_0}_{s_<} e^{-s/M_B^2} s \rho_{ii}(s) ds}{\int^{s_0}_{s_<} e^{-s/M_B^2} \rho_{ii}(s) ds} \, .
\label{eq:LSR}
\end{equation}
%

We use the spectral density $\rho_{11}(s)$ extracted from the current $\eta_{0^{+-};1}$ as an example to calculate the mass $M_1$ of the state $X_1$. As given in Eq.~(\ref{eq:LSR}), the mass $M_1$ depends on two free parameters: the Borel mass $M_B$ and the threshold value $s_0$. We consider three aspects to find their proper working regions: a) the OPE convergence, b) the one-pole-dominance assumption, and c) the dependence of the mass $M_1$ on these two parameters.

%
\begin{figure}[hbt]
\begin{center}
\includegraphics[width=0.47\textwidth]{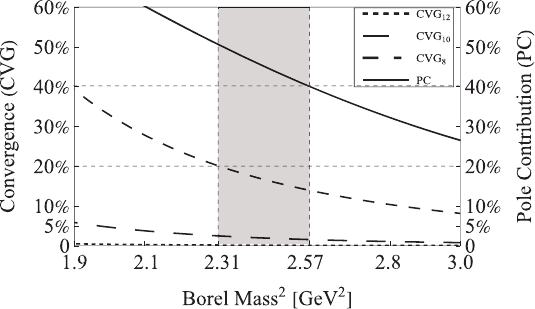}
\caption{CVG$_{8/10/12}$ and PC with respect to the Borel mass $M_B$. These curves are obtained using the current $\eta_{0^{+-};1}$ by setting $s_0 = 14.0$~GeV$^2$.}
\label{fig:cvgpole}
\end{center}
\end{figure}
%

Firstly, we consider the OPE convergence and require the $D=12/10/8$ terms to be less than 5\%/10\%/20\%, respectively:
\begin{eqnarray}
\mbox{CVG}_{12} &=& \left|\frac{ \Pi_{11}^{D=12}(\infty, M_B^2) }{ \Pi_{11}(\infty, M_B^2) }\right| \leq 5\% \, ,
\label{eq:cvg12}
\\
\mbox{CVG}_{10} &=& \left|\frac{ \Pi_{11}^{D=10}(\infty, M_B^2) }{ \Pi_{11}(\infty, M_B^2) }\right| \leq 10\% \, ,
\label{eq:cvg10}
\\
\mbox{CVG}_8 &=& \left|\frac{ \Pi_{11}^{D=8}(\infty, M_B^2) }{ \Pi_{11}(\infty, M_B^2) }\right| \leq 20\% \, .
\label{eq:cvg8}
\end{eqnarray}
These conditions demand the Borel mass to be larger than $M_B^2 \geq 2.31$~GeV$^2$, as depicted in Fig.~\ref{fig:cvgpole}.

Secondly, we consider the one-pole-dominance assumption and require the pole contribution to be larger than 40\%:
\begin{equation}
\mbox{PC} = \left|\frac{ \Pi_{11}(s_0, M_B^2) }{ \Pi_{11}(\infty, M_B^2) }\right| \geq 40\% \, .
\label{eq:pc}
\end{equation}
This condition demands the Borel mass to be smaller than $M_B^2 \leq 2.57$~GeV$^2$ when setting $s_0 = 14.0$~GeV$^2$, as depicted in Fig.~\ref{fig:cvgpole}.

Altogether we determine the Borel window to be $2.31$~GeV$^2 \leq M_B^2 \leq 2.57$~GeV$^2$ for $s_0 = 14.0$~GeV$^2$. We redo the same procedures and find that the Borel windows exist as long as $s_0 \geq s_0^{\rm min} = 12.5$~GeV$^2$. Accordingly, we demand the threshold value $s_0$ to be slightly larger and choose its working region to be $11.0$~GeV$^2 \leq s_0 \leq 17.0$~GeV$^2$.

\begin{figure*}[hbtp]
\begin{center}
\subfigure[]{\includegraphics[width=0.4\textwidth]{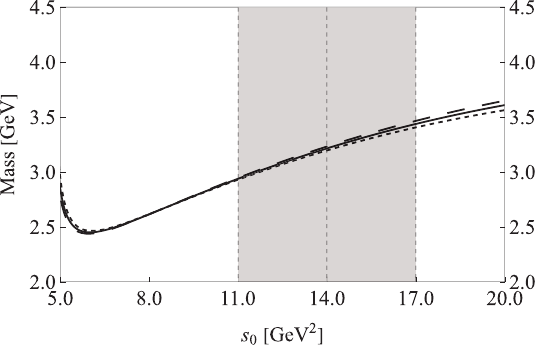}}
~~~~~~~~~~
\subfigure[]{\includegraphics[width=0.4\textwidth]{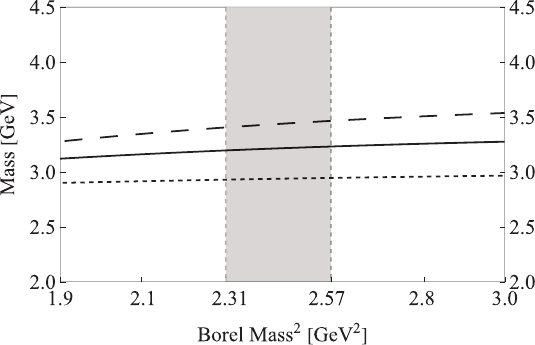}}
\caption{The mass $M_1$ of the state $X_1$ with respect to (a) the threshold value $s_0$ and (b) the Borel mass $M_B$. In the subfigure (a) the short-dashed/solid/long-dashed curves are obtained by setting $M_B^2 = 2.31/2.44/2.57$~GeV$^2$, respectively. In the subfigure (b) the short-dashed/solid/long-dashed curves are obtained by setting $s_0 = 11.0/14.0/17.0$~GeV$^2$, respectively. These curves are obtained using the spectral density $\rho_{11}(s)$ extracted from the current $\eta_{0^{+-};1}$.}
\label{fig:mass}
\end{center}
\end{figure*}

Thirdly, we consider the dependence of the mass $M_1$ on $M_B$ and $s_0$. As shown in Fig.~\ref{fig:mass}, the mass $M_1$ is stable against $M_B$ inside the Borel window $2.31$~GeV$^2 \leq M_B^2 \leq 2.57$~GeV$^2$, and its dependence on $s_0$ is acceptable inside the working region $11.0$~GeV$^2 \leq s_0 \leq 17.0$~GeV$^2$, where the mass $M_1$ is calculated to be
\begin{equation}
M_{0^{+-};1} = 3.21^{+0.23}_{-0.28}{\rm~GeV} \, .
\end{equation}
Its uncertainty comes from $M_B$ and $s_0$ as well as various QCD parameters given in Eqs.~(\ref{eq:condensates}). Note that the mass $M_1$ has a stability point at around $s_0 \sim 6.0$~GeV$^2$, as shown in Fig.~\ref{fig:mass}(a). However, there does not exist the Borel window at this energy point.

We apply the same method to study the other two $J^{PC} = 0^{+-}$ currents $\eta_{0^{+-};2/3}$ and the three $J^{PC} = 2^{+-}$ currents $\eta_{2^{+-};1/2/3}^{\beta_1 \beta_2}$. The obtained results are summarized in Table~\ref{tab:result}.

\begin{table*}[hpt]
\begin{center}
\renewcommand{\arraystretch}{1.6}
\caption{QCD sum rule results for the fully-strange tetraquark states with the exotic quantum numbers $J^{PC} = 0/2/4^{+-}$, extracted from the diquark-antidiquark currents $\eta_{0/2/4^{+-};1/2/3}^{\cdots}$ as well as their mixing currents $J_{0/2/4^{+-};1/2/3}^{\cdots}$. The results for the exotic quantum numbers $J^{PC} = 4^{+-}$ are taken from Ref.~\cite{Dong:2022otb}.}
\begin{tabular}{c|c|c|c|c|c}
\hline\hline
~~\multirow{2}{*}{Currents}~~ & ~$s_0^{min}$~ & \multicolumn{2}{c|}{Working Regions} & ~~\multirow{2}{*}{Pole~[\%]}~~ & ~~\multirow{2}{*}{Mass~[GeV]}~~
\\ \cline{3-4}
& ~~$[{\rm GeV}^2]$~~ & ~~$M_B^2~[{\rm GeV}^2]$~~ & ~~$s_0~[{\rm GeV}^2]$~~ &&
\\ \hline\hline
$\eta_{0^{+-};1}$        &  12.5   &  $2.31$--$2.57$   &  $14\pm3.0$  &  $40$--$51$  &  $3.21^{+0.23}_{-0.28}$
\\
$\eta_{0^{+-};2}$        &  21.2   &  $2.91$--$3.31$   &  $23\pm5.0$  &  $40$--$51$  &  $4.41^{+0.36}_{-0.31}$
\\
$\eta_{0^{+-};3}$        &  8.3   &  $1.13$--$2.21$   &  $14\pm3.0$  &  $40$--$94$  &  $3.20^{+0.19}_{-0.30}$
\\ \hline
$J_{0^{+-};1}$           &  8.3   &  $1.33$--$1.42$   &  $9\pm2.0$  &  $40$--$47$  &  $2.47^{+0.33}_{-0.44}$
\\
$J_{0^{+-};2}$           &  7.9   &  $1.41$--$1.62$   & $9\pm2.0$   &  $40$--$54$  &  $2.56^{+0.25}_{-0.35}$
\\
$J_{0^{+-};3}$           &  --     &  --               &  --          &  --          &  --
\\ \hline\hline
$\eta_{2^{+-};1}^{\beta_1\beta_2}$                          &  12.8   &  $2.11$--$2.32$   &  $14\pm3.0$  &  $40$--$50$  &  $3.27^{+0.23}_{-0.28}$
\\
$\eta_{2^{+-};2}^{\beta_1\beta_2}$        &  17.1   &  $2.35$--$2.78$   &  $19\pm4.0$  &  $40$--$55$  &  $3.98^{+0.29}_{-0.25}$
\\
$\eta_{2^{+-};3}^{\beta_1\beta_2}$        &  9.3   &  $1.15$--$2.14$   &  $14\pm3.0$  &  $40$--$91$  &  $3.27^{+0.19}_{-0.23}$
\\ \hline
$J_{2^{+-};1}^{\beta_1\beta_2}$           &  11.6   &  $1.97$--$2.19$   &  $13\pm3.0$  &  $40$--$51$  &  $3.07^{+0.25}_{-0.33}$
\\
$J_{2^{+-};2}^{\beta_1\beta_2}$           &  16.0   &  $2.44$--$2.82$   & $18\pm4.0$   &  $40$--$54$  &  $3.77^{+0.24}_{-0.30}$
\\
$J_{2^{+-};3}^{\beta_1\beta_2}$           &  17.0   &  $2.34$--$2.77$   & $19\pm4.0$   &  $40$--$55$  &    $3.99^{+0.29}_{-0.25}$
\\ \hline\hline
$\eta_{4^{+-};1}^{\alpha_1\alpha_2\alpha_3\alpha_4}$        &  14.6   &  $2.40$--$2.65$   &  $16\pm3.0$  &  $40$--$50$  &  $3.50^{+0.21}_{-0.25}$
\\
$\eta_{4^{+-};2}^{\alpha_1\alpha_2\alpha_3\alpha_4}$        &  19.2   &  $2.80$--$3.13$   &  $21\pm4.0$  &  $40$--$51$  &  $4.08^{+0.26}_{-0.31}$
\\
$\eta_{4^{+-};3}^{\alpha_1\alpha_2\alpha_3\alpha_4}$        &  11.0   &  $1.25$--$2.44$   &  $16\pm3.0$  &  $40$--$91$  &  $3.51^{+0.20}_{-0.20}$
\\ \hline
$J_{4^{+-};1}^{\alpha_1\alpha_2\alpha_3\alpha_4}$           &  10.1   &  $1.78$--$1.92$   &  $11\pm2.0$  &  $40$--$48$  &  $2.85^{+0.19}_{-0.22}$
\\
$J_{4^{+-};2}^{\alpha_1\alpha_2\alpha_3\alpha_4}$           &  19.1   &  $2.79$--$3.14$   & $21\pm4.0$   &  $40$--$51$  &  $4.08^{+0.26}_{-0.31}$
\\
$J_{4^{+-};3}^{\alpha_1\alpha_2\alpha_3\alpha_4}$           &  --     &  --               &  --          &  --          &  --
\\ \hline\hline
\end{tabular}
\label{tab:result}
\end{center}
\end{table*}

\subsection{Multi-channel analysis}
\label{sec:mixing}

\begin{figure}[hbt]
\begin{center}
\includegraphics[width=0.48\textwidth]{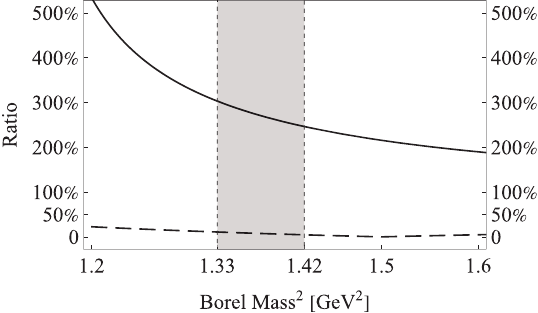}
\end{center}
\caption{The normalized off-diagonal correlation functions $\left|\Pi_{12}/\sqrt{\Pi_{11}\Pi_{22}}\right|$ (solid) and $\left|\Pi^\prime_{12}/\sqrt{\Pi^\prime_{11}\Pi^\prime_{22}}\right|$ (dashed) with respect to the Borel mass $M_B$. These curves are obtained using the three currents $\eta_{0^{+-};1/2/3}$ and their mixing currents $J_{0^{+-};1/2/3}$ when setting $s_0 = 9.0$~GeV$^2$.}
\label{fig:offdiagonal}
\end{figure}

In this subsection we perform the multi-channel analysis by taking into account the off-diagonal correlation functions that are actually non-zero, {\it i.e.}, $\rho_{ij}(s)|_{i \neq j} \neq 0$. To see how large they are, we choose $s_0 = 9.0$~GeV$^2$ and $M_B^2 = 1.50$~GeV$^2$ to obtain
\begin{equation}
\Pi_{ij}(s_0, M_B^2)
=
\left(\begin{array}{ccc}
8.29 & 13.59 & -2.83
\\
13.59 & -3.96 & -0.87
\\
-2.83 & -0.87 & 14.28
\end{array}\right) \times 10^{-6} {\rm~GeV}^{14} .
\end{equation}
This indicates that $\eta_{0^{+-};1}$ and $\eta_{0^{+-};2}$ are strongly correlated with each other, as depicted in Fig.~\ref{fig:offdiagonal}.

To diagonalize the $3\times3$ matrix $\rho_{ij}(s)$, we construct three mixing currents $J_{0^{+-};1/2/3}$:
\begin{equation}
\left(\begin{array}{c}
J_{0^{+-};1}
\\
J_{0^{+-};2}
\\
J_{0^{+-};3}
\end{array}\right)
=
\mathbb{T}_{0^{+-}}
\left(\begin{array}{c}
\eta_{0^{+-};1}
\\
\eta_{0^{+-};2}
\\
\eta_{0^{+-};3}
\end{array}\right)
\, ,
\label{eq:transition0}
\end{equation}
where $\mathbb{T}_{0^{+-}}$ is the transition matrix.

We use the method of operator product expansion to extract the spectral densities $\rho^\prime_{ij}(s)$ from the mixing currents $J_{0^{+-};1/2/3}$. After choosing
\begin{equation}
\mathbb{T}_{0^{+-}}
=
\left(\begin{array}{ccc}
-0.72 & -0.45 & 0.53
\\
0.54 & -0.84 & 0.03
\\
0.43 & 0.31 & 0.85
\end{array}\right) \, ,
\label{eq:matrix0}
\end{equation}
we obtain
\begin{equation}
\Pi^\prime_{ij}(s_0, M_B^2)
=
\left(\begin{array}{ccc}
12.52 & 0 & 0
\\
0 & 18.85 & 0
\\
0 & 0 & -12.77
\end{array}\right) \times 10^{-6} {\rm~GeV}^{14} ,
\label{eq:mixingpi4}
\end{equation}
at $s_0 = 9.0$~GeV$^2$ and $M_B^2 = 1.50$~GeV$^2$. Therefore, the off-diagonal terms of $\rho^\prime_{ij}(s)$ are negligible and the three mixing currents $J_{0^{+-};1/2/3}$ are nearly non-correlated around here, as depicted in Fig.~\ref{fig:offdiagonal}. Besides, Eq.~(\ref{eq:mixingpi4}) indicates that the QCD sum rule result extracted from $J_{0^{+-};3}$ is non-physical around here due to its negative correlation function.

Since the off-diagonal terms of $\rho^\prime_{ij}(s)$ are negligible around $s_0 = 9.0$~GeV$^2$ and $M_B^2 = 1.50$~GeV$^2$, the procedures used in the previous subsection can be applied to study the three mixing currents $J_{0^{+-};1/2/3}$. We summarize the obtained results in Table~\ref{tab:result}. Especially, the mass extracted from the current $J_{0^{+-};1}$ is significantly reduced to be
\begin{equation}
M^\prime_{0^{+-};1} = 2.47^{+0.33}_{-0.44} {\rm~GeV} \, .
\end{equation}

Similarly, we can investigate the three $J^{PC} = 2^{+-}$ currents $\eta_{2^{+-};1/2/3}^{\beta_1 \beta_2}$. We construct three mixing currents $J_{2^{+-};1/2/3}^{\beta_1 \beta_2}$ that are nearly non-correlated at around $s_0 = 13.0$~GeV$^2$ and $M_B^2 = 2.0$~GeV$^2$:
\begin{equation}
\left(\begin{array}{c}
J_{2^{+-};1}^{\cdots}
\\
J_{2^{+-};2}^{\cdots}
\\
J_{2^{+-};3}^{\cdots}
\end{array}\right)
=
\mathbb{T}_{2^{+-}}
\left(\begin{array}{c}
\eta_{2^{+-};1}^{\cdots}
\\
\eta_{2^{+-};2}^{\cdots}
\\
\eta_{2^{+-};3}^{\cdots}
\end{array}\right)
\, ,
\label{eq:transition4}
\end{equation}
where
\begin{equation}
\mathbb{T}_{2^{+-}}
=
\left(\begin{array}{ccc}
-0.72 &0.06  & 0.69
\\
-0.68  &0.15 & -0.72
\\
-0.15 &-0.99 & -0.07
\end{array}\right) \, .
\label{eq:matrix4}
\end{equation}
We apply the QCD sum rule method to study the mixing currents $J_{2^{+-};1/2/3}^{\beta_1 \beta_2}$, and the obtained results are summarized in Table~\ref{tab:result}. Especially, the mass extracted from the current $J_{2^{+-};1}^{\beta_1 \beta_2}$ is the lowest:
\begin{equation}
M^\prime_{2^{+-};1} = 3.07^{+0.25}_{-0.33} {\rm~GeV} \, .
\end{equation}

For completeness, we also summarize in Table~\ref{tab:result} the QCD sum rule results obtained in Ref.~\cite{Dong:2022otb} using the three $J^{PC} = 4^{+-}$ currents $\eta_{4^{+-};1/2/3}^{\alpha_1\alpha_2\alpha_3\alpha_4}$ as well as their mixing currents $J_{4^{+-};1/2/3}^{\alpha_1\alpha_2\alpha_3\alpha_4}$ that are nearly non-correlated at around $s_0 = 11.0$~GeV$^2$ and $M_B^2 = 1.85$~GeV$^2$:
\begin{equation}
\left(\begin{array}{c}
J_{4^{+-};1}^{\cdots}
\\
J_{4^{+-};2}^{\cdots}
\\
J_{4^{+-};3}^{\cdots}
\end{array}\right)
=
\mathbb{T}_{4^{+-}}
\left(\begin{array}{c}
\eta_{4^{+-};1}^{\cdots}
\\
\eta_{4^{+-};2}^{\cdots}
\\
\eta_{4^{+-};3}^{\cdots}
\end{array}\right)
\, ,
\label{eq:transition4}
\end{equation}
where
\begin{equation}
\mathbb{T}_{4^{+-}}
=
\left(\begin{array}{ccc}
0.72 & -0.06 & -0.69
\\
0.14 & 0.99 & 0.05
\\
0.68 & -0.13 & 0.72
\end{array}\right) \, .
\label{eq:matrix4}
\end{equation}

\section{Summary and Discussions}
\label{sec:summary}

In this paper we apply the QCD sum rule method to study the fully-strange tetraquark states with the exotic quantum numbers $J^{PC} = 0^{+-}$ and $2^{+-}$. We explicitly add the covariant derivative operator to construct some diquark-antidiquark interpolating currents, and apply the method of operator product expansion to calculate both their diagonal and off-diagonal correlation functions. Based on the obtained results, we construct some mixing currents that are nearly non-correlated.

We use both the diquark-antidiquark currents and their mixing currents to perform QCD sum rule analyses. The obtained results are summarized in Table~\ref{tab:result}. Especially, we use the mixing currents $J_{0^{+-};1}$ and $J_{2^{+-};1}^{\beta_1 \beta_2}$ to derive the masses of the lowest-lying $J^{PC} = 0^{+-}$ and $2^{+-}$ states to be
\begin{eqnarray} M_{0^{+-}} &=& 2.47^{+0.33}_{-0.44}{\rm~GeV} \, ,
\\[2mm]  M_{2^{+-}} &=& 3.07^{+0.25}_{-0.33}{\rm~GeV} \, .
\end{eqnarray}

In this paper we also construct some fully-strange meson-meson currents of $J^{PC} = 0^{+-}$ and $2^{+-}$, which are related to the diquark-antidiquark currents through the Fierz rearrangement. We can use these meson-meson currents and their mixing currents to perform QCD sum rule analyses. The results extracted from these mixing currents are the same.

We can apply Eqs.~(\ref{eq:fierz}) to transform the mixing currents $J_{0^{+-};1}$ and $J_{2^{+-};1}^{\beta_1 \beta_2}$ to be
\begin{eqnarray}
\label{eq:decay0}
J_{0^{+-};1}\!\!\! &=& \! \! -2.7~\xi_{0^{+-};1} - 1.6~\xi_{0^{+-};2} + 2.3~\xi_{0^{+-};3} \, ,
\\ J_{2^{+-};1}^{\cdots} \!\!\!&=&\!\! -4.3~\xi_{2^{+-};1}^{\cdots} - 1.2~\xi_{2^{+-};2}^{\cdots} + 1.3~\xi_{2^{+-};3}^{\cdots} \, . \label{eq:decay2}
\end{eqnarray}
For completeness, we compare these two combinations to the mixing current
\begin{equation}
\label{eq:decay4}
J_{4^{+-};1}^{\cdots} = 4.3~\xi_{4^{+-};1}^{\cdots} + 1.2~\xi_{4^{+-};2}^{\cdots} - 1.3~\xi_{4^{+-};3}^{\cdots} \, ,
\end{equation}
which was used in Ref.~\cite{Dong:2022otb} to derive the mass of the lowest-lying $J^{PC} = 4^{+-}$ state to be
\begin{eqnarray}
 M_{4^{+-}} &=& 2.85^{+0.19}_{-0.22}{\rm~GeV} \, .
\end{eqnarray}

The Fierz identity given in Eq.~(\ref{eq:decay0}) indicates that the lowest-lying $J^{PC} = 0^{+-}$ state decays into the $P$-wave $\phi(1020) f_0(1710)(\to \phi K \bar K/\phi\pi \pi)$ channel through the meson-meson current $\xi_{0^{+-},1}$, and the Fierz identity given in Eq.~(\ref{eq:decay2}) indicates that the lowest-lying $J^{PC} = 2^{+-}$ state decays into the $P$-wave $\phi(1020) f_0(1710)$ and $\phi(1020) f_2^\prime(1525)$ channels through the meson-meson current $\xi_{2^{+-},1}^{\cdots}$. Accordingly, we propose to search for them in the $X_{0^{+-}} \to \phi(1020) f_0(1710) \to \phi K \bar K/\phi\pi \pi$ and $X_{2^{+-}} \to \phi(1020) f_0(1710)/\phi(1020) f_2^\prime(1525) \to \phi K \bar K/\phi\pi \pi$ decay processes in the future Belle-II, BESIII, COMPASS, GlueX, J-PARC and PANDA experiments.

\section*{Acknowledgments}
%

This project is supported by
the National Natural Science Foundation of China under Grant No.12075019,
the Japanese Grant-in-Aid for Scientific Research under Grant Nos. 21H04478 and 18H05407,
the Jiangsu Provincial Double-Innovation Program under Grant No.~JSSCRC2021488,
and
the Fundamental Research Funds for the Central Universities.

\bibliographystyle{elsarticle-num}
\bibliography{ref}

\end{document}